\documentclass[twocolumn]{iopjournal_twocolumn}

\usepackage{ulem}
\usepackage{amsmath}
\usepackage{siunitx}
\begin{document}

\articletype{Paper} 

\title{Resonant Pitch-Angle Scattering Of Runaway-Electrons by Externally-launched Helicon Waves in the DIII-D Tokamak}

\author{
H.~Choudhury$^{1}$\orcid{0000-0000-0000-0000},
J.~Lestz$^{3,5}$\orcid{0000-0000-0000-0000},
C.~Paz-Soldan$^{1,*}$\orcid{0000-0000-0000-0000},
A.~Battey$^{1,2}$\orcid{0000-0000-0000-0000},
N.~Leuthold$^{1}$\orcid{0000-0000-0000-0000},
A.~Lvovskiy$^{3}$\orcid{0000-0000-0000-0000},
C.~Marini$^{4}$\orcid{0000-0000-0000-0000},
J.~Barr$^{3}$\orcid{0000-0000-0000-0000},
W.~Heidbrink$^{5}$\orcid{0000-0000-0000-0000},
D.~Spong$^{6}$\orcid{0000-0000-0000-0000},
S.~Tang$^{3}$\orcid{0000-0000-0000-0000},
B.~Van~Compernolle$^{3}$\orcid{0000-0000-0000-0000},
Q.~Zhang$^{7}$\orcid{0000-0000-0000-0000},
Y.~Zhang$^{7}$\orcid{0000-0000-0000-0000},
and X.~Tang$^{7}$\orcid{0000-0000-0000-0000}
}

\affil{$^{1}$Columbia University, New York, New York 10027, USA}
\affil{$^{2}$EPFL, Swiss Plasma Center (SPC), Lausanne CH-1015, Switzerland}
\affil{$^{3}$General Atomics, San Diego, California 92186, USA}
\affil{$^{4}$University of California San Diego, La Jolla, California 92093, USA}
\affil{$^{5}$University of California, Irvine, California 92697, USA}
\affil{$^{6}$Oak Ridge National Laboratory, Oak Ridge, Tennessee 37831, USA}
\affil{$^{7}$Los Alamos National Laboratory, Los Alamos, New Mexico 87545, USA}
\affil{$^*$Author to whom any correspondence should be addressed.}
 
\email{carlos.pazsoldan@columbia.edu}  

\pagebreak
\begin{abstract}

Resonant wave-particle interactions between externally launched helicon waves (also known as whistler waves) and runaway electrons (REs) have been demonstrated on the DIII-D tokamak. In this work we extend the initial results reported in \textit{Choudhury, H. et al. Phys. Rev. Lett. 136, 025101 (2026)} by exploring the effects of antenna alignment with the edge magnetic field, toroidal wave propagation direction, and coupled power on RE scattering in the quiescent RE experimental scenario. Two distinct experimental configurations have been investigated: one in which the antenna aligns well with the edge background magnetic field, known as the ideal antenna configuration, and one with misalignment, known as the non-ideal case. Previously, it had been found that helicon power in the ideal antenna configuration prevented RE growth despite the normalized toroidal electric field remaining high enough to drive exponential RE growth in the absence of helicon power. In this paper, we show that scattering via the normal Doppler resonance ($n=1$) effectively limits the growth of the RE population in both the ideal and non-ideal antenna configurations, with evidence of a power threshold in the latter case. In contrast, launching waves that favour the anomalous Doppler resonance ($n=-1$) is observed to \textit{enhance} rather than reduce the RE population. In addition, fast magnetic measurements reveal rising-tones in the \SIrange{30}{60}{\mega\hertz} range during helicon-off periods, which are not observed prior to helicon power. Finally, the challenges of using launched helicon waves to scatter post-disruption RE beams are discussed. Collisional damping and a large vacuum gap between the plasma and antenna on the outboard side present significant obstacles to helicon waves propagating into the plasma core.
\end{abstract}
\pagebreak
 
\pagebreak

\section{Introduction}

For electrons traveling faster than the thermal speed in a plasma, the collisional drag force is equal to 
\begin{equation}
    F_{drag} = m \nu_{e e} v = \frac{n_e e^4 \ln \Lambda }{4 \pi \varepsilon_0^2 m_e v^2}
    \label{eq:dragforce}
\end{equation}
where $\nu_{ee}=n_e e^4 \ln \Lambda /\left(4 \pi \varepsilon_0^2 m_e^2 v^3\right)$ is the electron-electron collision frequency, $n_e$ is the electron density, $e$ is the elementary charge, $m_e$ is the electron mass, $v$ is the electron velocity, $\varepsilon_0$ is the permittivity of free space, and $\ln \Lambda$ is the Coulomb logarithm \cite{connorrelativisticlimitationsrunaway1975,dreicerelectronionrunaway1960,granetzitpajointexperiment2014}. Because this force \textit{decreases} with increasing velocity, an electron population subject to a sufficiently strong electric field can run away to relativistic energies, a process known as Dreicer generation \cite{dreicerelectronionrunaway1960}. For particles with significant perpendicular energy or traveling on curved field lines, radiative losses such as synchrotron become increasingly significant at higher energies \cite{breizmanphysicsrunawayelectrons2019}. 
\par
Tokamaks drive a toroidal plasma current using an inductively applied electric field, which, under certain conditions, can create REs. For example, following an unintended disruption, if a significant fraction of the plasma current is carried by REs instead of the thermal electrons, the highly localized energy deposition from this RE beam onto plasma-facing components could cause severe damage \cite{boozerpivotalissuesrelativistic2018a}. REs, therefore, present a unique problem for future high-current tokamaks such as ITER \cite{boozerrunawayelectronsiter2017a}. The electric field threshold below which no runaway electrons (REs) are generated is known as the Connor-Hastie critical electric field, $E_{CH}$ \cite{connorrelativisticlimitationsrunaway1975}:
\begin{equation}
    E_{CH} = \frac{n_e e^3 \ln \Lambda}{4 \pi \varepsilon_0^2 m_e c^2}
    \label{eq:ECH}
\end{equation}
where $c$ is the speed of light in vacuum. The derivation of $E_{CH}$ in Ref. \cite{connorrelativisticlimitationsrunaway1975} accounts for the relativistic nature of these energetic particles, but does not consider the possibility of loss mechanisms other than collisional drag on electrons. It has been found experimentally that some multiple of $E_{CH}$ is required to generate REs in tokamaks \cite{granetzitpajointexperiment2014}, likely due to additional loss mechanisms such as synchrotron radiation, radiative losses, and kinetic instabilities \cite{liueffectskineticinstabilities2018,paz-soldangrowthdecayrunaway2014,deckerexpulsionrunawayelectrons2024}. 
\par

\par
In addition to Dreicer generation, there are other processes that can generate REs in tokamaks, such as hot-tail generation during rapid cooling events and Compton scattering of high-energy photons from neutron-activated first wall components in future devices \cite{breizmanphysicsrunawayelectrons2019}. A distinct secondary process, known as avalanche growth, involves a RE striking a thermal electron, imparting enough energy to it to cause it to runaway, whilst remaining a RE itself \cite{rosenbluththeoryavalancherunaway1997}. The avalanche process is thought to be particularly problematic for future high-current devices, though we note here that for the experiments presented in this work, it is the primary Dreicer process that dominates RE generation.
\par
REs can arise in three broad experimental categories in tokamaks, distinguished roughly by whether the plasma current is ramping up, steady, or crashing, as in a current quench. These three categories are start-up, quiescent (flat-top), and disruption. 
\par
In tokamak start-up scenarios, REs can be generated as the plasma current is ramping up and the plasma density is increasing as gas is being puffed in. These scenarios can have high normalized electric fields, $E/E_{CH}$, where $E$ is the toroidal electric field inside the plasma. If the density is not increased fast enough, the RE population can grow and `overtake' the thermal electrons to become the dominant current carrier \cite{hoppeRunawayElectronGeneration2022}.
\par
In a low-density ohmic plasma, which can require careful error-field correction in a medium-sized device like DIII-D \cite{pironerrorfielddetection2024,paz-soldannondisruptiveerrorfield2022,paz-soldanspectralbasisoptimal2014,scovillelockedmodesdiiid1991}, it is possible for a measurable yet benign trace population of REs to be generated during the phase of constant current in tokamaks. If the error-field correction is not done accurately, these plasmas develop fatal locked-modes. On DIII-D, this is known as the Quiescent Runaway Electron Scenario (QRE) \cite{paz-soldangrowthdecayrunaway2014,spongfirstdirectobservation2018,heidbrinklowfrequencywhistlerwaves2018}. In this scenario, the normalized electric field is typically in the range $5 < E/E_{CH} < 15$ and the RE current is a small fraction of the total plasma current (typically of the order of one percent). All the experimental data presented in this work are from QRE discharges on DIII-D because QRE plasmas are long-lived, generally steady state, benign, reproducible plasmas that enable wave-particle interactions to be studied.
\par
Finally, during disruptions, the plasma rapidly cools and the plasma resistivity increases dramatically, leading to a large toroidal electric field that can generate REs which carry the bulk of the pre-disruption plasma current. If control of this beam of REs is lost and it strikes plasma-facing components with significant current, it can cause severe damage, such as exposing cooling channels. Repairing such components is all the more challenging in future devices that use deuterium-tritium fuel cycles, due to the neutron activation of materials \cite{boozerPlasmaSteeringAvoid2021}. In addition to concerns about REs, disruptions can cause large electromagnetic stresses on the vacuum vessel and associated structures through so-called Halo currents \cite{boozertokamakhalocurrents2013}. It was said well in Ref \cite{boozerPlasmaSteeringAvoid2021}: ``Steering a tokamak to avoid disruptions resembles driving a car at high speed through a dense fog."
\par
Several strategies have been proposed to mitigate the threat of REs in tokamaks. Since the drag on electrons increases with density, as seen from Eq. \ref{eq:dragforce}, increasing the plasma density can help suppress RE generation. Massive gas injection (MGI) and shattered pellet injection (SPI) are two techniques that have been developed to rapidly increase the plasma density, but have so far had mixed results in terms of RE suppression \cite{shirakidissipationpostdisruptionrunaway2018,reuxrunawayelectronbeam2015}. One technique, which is intended to be a passive technique, i.e. one that does not require sophisticated control systems and turns on when needed, is the Runaway Electron Mitigation Coil (REMC). It is still to be experimentally tested but the idea is as follows: during a current quench, the current induced in an REMC generates non-axisymmetric magnetic perturbations which deconfine suprathermal electrons before they can reach relativistic energies \cite{boozertwobeneficialnonaxisymmetric2011,smithpassiverunawayelectron2013}. Other strategies for RE prevention and mitigation include resonant magnetic perturbations \cite{lehnensuppressionrunawayelectrons2008}, inducing MHD instabilities to deconfine REs \cite{paz-soldannovelpathrunaway2021,reuxdemonstrationsafetermination2021}, and the use of electron-cyclotron heating to increase RE transport \cite{deckerexpulsionrunawayelectrons2024}. In general, RE mitigation strategies can be broadly grouped into those that increase collisional damping of the REs, such as MGI and SPI, or collisionless methods that aim to deconfine REs directly, like the REMC or inducing MHD instabilities (the latter technique is also known as benign termination). One interesting observation of a passive mitigation technique was the appearance of compressional Alfven modes in post-disruption current quench plasmas on DIII-D that induced RE losses \cite{lvovskiyparametricstudyalfv2022,liucompressionalalfveneigenmodes2021,liurolekineticinstability2018}. The losses increased with decreasing magnetic field because of increases in energy of the kinetic instabilities, and no RE-beam was found to be formed for fields of less than \SI{1.8}{T}, which was explained by the presence of strong instabilities. Relying on the RE distribution itself though to excite waves which deconfine REs may not be the most practical mitigation strategy since one cannot guarantee the post-disruption RE distribution will excite the required modes.
\par
In this paper we focus on the use of resonant wave-particle interactions to pitch-angle scatter REs to higher perpendicular energies and thus increase their synchrotron damping and hence decrease their total energy. Though pitch-angle scattering has been observed by \textit{RE-driven} whistler waves \cite{choudhurydetailedcharacterizationrunaway2025}, the waves here are externally launched from an antenna on the outer wall of the device. The waves are right-handed electromagnetic radio-frequency helicon waves, also known as whistler waves or the fast wave in the lower hybrid range \cite{pinskerwhistlersheliconslower2015}. There is an approximate naming convention that the experimental team at DIII-D uses: when the waves are launched from the helicon antenna they are called helicon waves, when the waves are driven by REs they are called whistler waves \cite{spongfirstdirectobservation2018,heidbrinklowfrequencywhistlerwaves2018}, and are referred to as the fast wave when needing to be contrasted with the slow wave, which is also referred to as the lower hybrid wave \cite{pinskerwhistlersheliconslower2015}. The theoretical basis for this technique in tokamaks was first proposed by Guo et al in Ref. \cite{guocontrolrunawayelectron2018} and experimentally demonstrated for the first time in the DIII-D QRE scenario in Ref. \cite{choudhuryFirstDemonstrationResonant2026}. Here we describe in greater detail the experiment whose main results were described in Ref. \cite{choudhuryFirstDemonstrationResonant2026}. 
\par
It is worth noting that resonant pitch-angle scattering using externally launched waves is a case of momentum-space engineering, in that the aim is to manipulate the RE vortex in momentum space directly \cite{guophasespacedynamicsrunaway2017,paz-soldanresolvingrunawayelectron2018}. Herein lies this method's main advantage: specific parts of momentum space can be targeted, depending on wave and plasma parameters. For example, a system could conceivably be designed to scatter REs of a specific energy, \SI{10}{MeV} say, since knowing the wave's wavevector and frequency and the plasma's magnetic field fixes the location of the resonance in momentum space. Other techniques might not limit the maximum energy of REs in this way, and might aim to reduce the energies of all REs or the bulk of the RE population. The main challenges with this technique concern the production and propagation of fast waves: helicon antenna technology is much less developed than, say, ECH technology, and fast waves are evanescent in vacuum where ECH is not. All of these aspects of helicon scattering of REs are described in detail in this paper.

\subsection{RE-wave interactions in space}
The idea of launching right-handed waves to scatter energetic electrons has been considered extensively in the context of Earth's radiation belts, in an idea sometimes called `radiation belt remediation' \cite{inancontrolledprecipitationradiation2003,kulkarniEnergeticElectronPrecipitation2008}. The Van Allen belts are two regions of energetic particles trapped by the Earth's magnetic field \cite{vanallengeomagneticallytrappedcorpuscular1959}. The population of energetic electrons in the Van Allen belts can vary depending on geomagnetic activity and high-altitude nuclear tests \cite{vanallengeomagneticallytrappedcorpuscular1959,vanallenobservationhighintensity1958}. Understanding these energetic particles is important because of their potential to damage spacecraft \cite{bakerhowcopespace2002,bakeroccurrenceoperationalanomalies2000}. Precipitation of the energetic electrons by whistler waves could be achieved using ground-based antennas or in-situ spacecraft. In addition, the idea of using left-handed electromagnetic ion cyclotron (EMIC) waves in space to scatter energetic protons has also been considered, but the engineering of such a system is much more challenging \cite{desoria-santacruzpichcontrolledprecipitationenergetic2014}.

\subsection{RE-driven waves}
In both space and tokamak plasmas, through the anisotropy of their distribution function, the energetic electrons are able to excite electromagnetic waves. In tokamaks, several kinetic instabilities have been experimentally predicted \cite{aleynikovstabilityanalysisrunawaydriven2015,castaldostabilityanalysisplasma2024} and observed, such as whistler waves during QRE shots on DIII-D \cite{spongfirstdirectobservation2018,heidbrinklowfrequencywhistlerwaves2018,choudhurydetailedcharacterizationrunaway2025}, compressional alfven waves during the current quench of post-disruption shots on DIII-D \cite{lvovskiyparametricstudyalfv2022,liucompressionalalfveneigenmodes2021,liurolekineticinstability2018}, chirping instabilities on DIII-D \cite{lvovskiyobservationrapidfrequency2019}, and, finally, lower hybrid waves on the FTU tokamak \cite{binfirstintrashotobservation2022}.

\section{Method}

\subsection{The Quiescent RE Experimental Scenario}
\label{sec:QREscenario}
In the QRE experimental scenario, low densities ($\sim \SI{5e18}{\per\cubic\metre}$) are used to achieve elevated normalized electric fields ($5 < E/E_{CH} < 15$) during the flat-top phase of the plasma current. The plasma current is typically $\sim \SI{0.8}{MA}$ and the toroidal magnetic field is either \SI{1.4}{T} or \SI{1.6}{T} depending on the configuration. There are typically two phases of a QRE shot: the growth phase, in which the RE population grows exponentially, and a steady-state phase, which is reached when gas is puffed to increase the density and stabilize the RE population \cite{spongfirstdirectobservation2018,heidbrinklowfrequencywhistlerwaves2018,paz-soldanrecentdiiidadvances2019}. QRE shots can last for up to \SI{10}{s} on DIII-D with the gas puff typically happening between \SI{4}{s} and \SI{6}{s}.
\par
All the experimental data presented in this paper, however, \textit{is only from the growth phase of QRE shots}. This was because the error-field correction proved challenging and it was not always possible to confine the plasma for long-enough to puff in gas and reach the steady-state phase. The shots presented here typically lasted \SIrange{6}{8}{s} with helicon power applied typically between \SIrange{3}{5}{s}. During the growth phase, absent any external effects, there is an exponential increase in the RE population and the energy spectrum of the REs shifts to higher energies.
\par
The toroidal electric field on the DIII-D tokamak, which is the focus of this work, is calculated as $E = V_{loop}/(2 \pi R)$, where $V_{loop}$ is the toroidal loop voltage measured and $R$ the major radius. The loop voltage is measured using an axisymmetric single-turn conducting flux loop on the horizontal plane \cite{straitMagneticDiagnosticSystem2006}. In these experiments, the plasma current is steady state and so the loop voltage gives an accurate measure of the electric field inside the plasma.
\par

\subsection{The Helicon Antenna and its alignment}
The helicon antenna on DIII-D is a 30-module megawatt-level traveling wave (comb-line) antenna that launches helicon waves in a \SI{10}{MHz}
band centered at a frequency of \SI{476}{MHz} \cite{nagyhighpowerhelicon2017,compernollehighpowerheliconprogram2021}. The antenna was originally installed and optimised to study radio-frequency current drive \cite{praterapplicationveryhigh2014}, though its wave parameters fortuitously overlap with those needed to resonantly interact with REs \cite{guocontrolrunawayelectron2018}. 
\par
The antenna, positioned on DIII-D's outer-wall at a major radius of \SI{2.27}{m}, spans roughly \SI{40}{\degree} in the toroidal direction. The plasma-facing side of the modules of the antenna is protected by a set of Faraday screen elements. The modules are tilted at \SI{14}{\degree} so the Faraday screen elements align with the edge background magnetic field that would be present during the target high-beta current-drive scenario. This scenario utilizes what is called `reversed $ B_T$' and `standard $ I_p$'. The `standard' and `reversed' terminology, which is admittedly confusing, is clarified in Table \ref{tab:stdrev_terminology}.
\begin{table}[h]
\centering
\caption{Direction as viewed from above the tokamak}
\begin{tabular}{c|c|c}
    &  $ I_p$ Direction & $ B_T$ Direction \\
\hline
`Standard' & Anti-clockwise & Clockwise \\
\hline  
`Reversed' & Clockwise & Anti-clockwise 
\label{tab:stdrev_terminology}
\end{tabular}
\end{table}

\par
\begin{figure*}[h]
    \centering    \includegraphics[width=\textwidth]{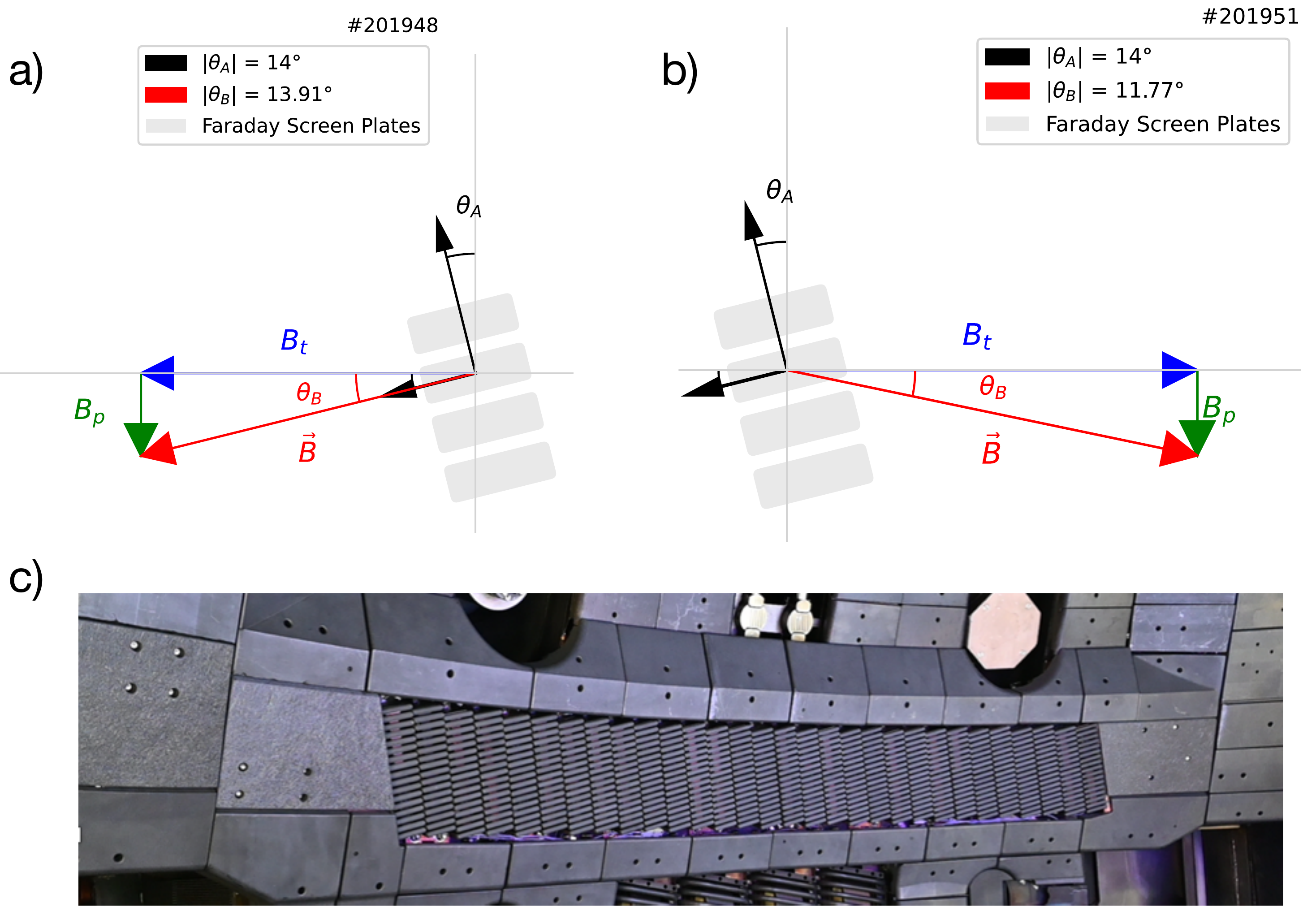}
    \caption{The alignment of the background edge magnetic field with the antenna's Faraday screen elements in the ideal case (a) or non-ideal case (b). The background field was obtained from the EFIT equilibria for both shots. Shown in (c) is a photograph of the antenna mounted in vessel. In the ideal case, the edge magnetic field is as well aligned as possible with the Faraday screen elements. This means the longitudinal component of the wave electric field is expected to be shorted and minimized, so that most of the power is put into the fast wave. To understand the schematics in (a) and (b), imagine looking at the antenna on the outer wall, the arrows for the magnetic field are relatively sized, to show the angle the edge $\vec{B}$ field makes. The Faraday screen elements are just schematic and not to scale, though the \SI{14}{\degree} tilt is accurate.  }
    \label{fig:fieldalignment}
\end{figure*}

Good alignment of the Faraday screen with the edge magnetic field direction results in less power from the antenna being directly excited into the slow wave, also known as the lower hybrid wave, which is an undesirable wave for this experiment since it has predominantly a longitudinal wave electric field (longitudinal here meaning parallel to the background magnetic field). We will, therefore, refer to this configuration in which the antenna is well aligned with the background magnetic field at the face of the antenna as the `ideal' configuration. The direction of the plasma current, and hence the runaway electrons, was not varied in these experiments, and so was throughout in the standard direction (i.e. $I_p$ was clockwise as viewed from above, and the bulk RE velocity was anti-clockwise). The magnetic field direction \textit{was} changed during the experiment, and so both the ideal and non-ideal configurations were explored. The alignment of the background magnetic field with the Faraday screen elements in both the ideal and non-ideal cases is illustrated schematically in Fig. \ref{fig:fieldalignment}, alongside a photograph of the antenna in the DIII-D vessel. In panel (a) of Fig. \ref{fig:fieldalignment}, the ideal case is shown in which the background edge magnetic field almost perfectly aligns with the Faraday screen elements. In panel (b), the non-ideal case is shown in which the background magnetic field is misaligned with respect to the Faraday screen elements by roughly \SI{26}{\degree} ($= |\theta_A|+|\theta_B| = 14^{\circ} + 11.77^{\circ}$). It is worth noting that the maximal misalignment between the Faraday screen elements and background edge magnetic field is \SI{90}{\degree}, since if the edge magnetic field were anti-aligned with the Faraday screen elements, the longitudinal electric field could still be shorted. To put it another way, let $\vec{f}$ be a vector aligned with the Faraday screen elements and $\vec{B}_{0,\textrm{edge}}$ be the field at the edge, then the ideal case is whenever $\vec{f} \times \vec{B}_{0,\textrm{edge}}=0$ and the non-ideal case is whenever $\vec{f} \cdot \vec{B}_{0,\textrm{edge}}=0$.
\par

A summary table of the number of shots in each experimental configurations explored is given in Table \ref{tab:experimental_configurations}.
\begin{table}[h]
\centering
\caption{Number of shots in each experimental configuration}  
\begin{tabular}{l|c|c}
    & Ideal & Non-ideal \\
\hline
Standard $ I_p$ & 15 shots & 3 shots 
\end{tabular}
\label{tab:experimental_configurations}
\end{table}
\par
It is worth noting that impurities from the Faraday screen elements, which are made of a boron-carbide-coated TZM molybdenum alloy, have not been observed to be a significant source of plasma contamination during helicon operation \cite{pinskerfirsthighpowerhelicon2024}.
\subsection{Helicon Power}
\label{subsec:heliconpower}
Power for the helicon antenna is provided by a klystron with a maximum output of \SI{1.2}{MW}. The 
\SI{476}{MHz} klystron, with a \SI{1}{dB} bandwidth of \SI{\pm1.5}{MHz}, was originally used at SLAC \cite{jongewaard1996operating}, before being transferred to DIII-D, where it was installed and integrated with a new control system \cite{compernollehighpowerheliconprogram2021}. The power from the klystron, which is measured on DIII-D, is attenuated by losses in the vacuum transmission line between the klystron and antenna. There are an array of probes inside the antenna which measure how much power leaves the antenna towards the plasma. Not all power leaves the antenna since it is of the comb-line type and some power makes it to the end of the antenna where it is deposited on a collector. In this paper, this power leaving the antenna is referred to as the `coupled power' or the `helicon power'. It is important to note that since the waves are evanescent in vacuum, the coupled power is an upper limit on how much power actually crosses the open-field line region between antenna and plasma (which can be either vacuum or filled with a low-density diffuse plasma) and makes it into the plasma. Great care was taken during this experiment to optimize the vacuum gap between plasma and antenna on the outer wall to ensure helicon power crossed the vacuum and made it to the core. The outboard gap was reduced to \SI{4.5}{cm} from \SI{8.5}{cm} for this experiment for ideal configuration shots. Significant helicon power deposition has been observed in other DIII-D shots with outboard gaps within the range \SIrange{5}{8}{cm} \cite{pinskerfirsthighpowerhelicon2024}. 
\par
A better metric than the outboard gap, though harder to measure because of difficulties measuring densities in the far scrape-off layer, is the distance between antenna and the fast-wave lower cut-off density. There is a trade-off to minimizing such a distance, however, since if the plasma is too close to the antenna, power will be launched from the first few antenna modules, and the rest will launch no power at all. This would result in a poorly defined $n_\parallel$ spectrum. This could happen even when the antenna-plasma distance is large enough for safe operation of the antenna. A poorly defined $n_\parallel$ spectrum makes it harder to pitch-angle scatter REs at specific resonant energies. It is therefore desirable to minimize the gap as much as possible subject to the constraint that all the power is leaving the antenna through as many of the antenna modules as possible, to achieve a well-defined $n_\parallel$ spectrum. In practice, because the outboard gap is not perfectly correlated with this better metric, it takes a few experimental shots to achieve the right plasma-antenna separation.
\par
Finally, the fraction of power launched directly into the fast wave versus the slow wave cannot be experimentally determined, but since, particularly in the ideal configuration, the alignment and gap are conducive to fast wave excitation and we observe evidence of wave-particle resonance possible only between the fast wave and REs, and the antenna can certainly launch fast waves \cite{pinskerfirsthighpowerhelicon2024}, we conclude that a sufficient fraction of the coupled power is launched into the fast wave.

\subsection{Fast wave propagation}
The refractive index of the wave launched from the antenna, set by the design of the modules, has a value of $n_{\parallel, antenna} = k_\parallel c/\omega = 3$, which corresponds to a parallel wavelength $\lambda_\parallel = \SI{21}{cm}$ at \SI{476}{MHz} \cite{compernollehighpowerheliconprogram2021}.
\begin{figure}
    \centering
    \includegraphics[width=\columnwidth]{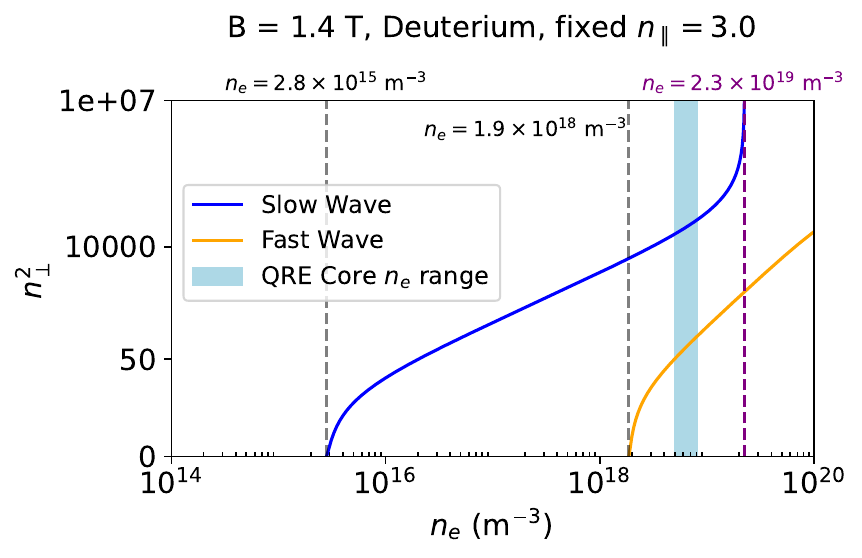}
    \caption{The cold-plasma dispersion relations for a purely deuterium plasma in a \SI{1.4}{T} field with wave $n_\parallel=3$, its value at the antenna.}
    \label{fig:nperpsquaredvsdensity}
\end{figure}
The cold-plasma wave dispersion relation is shown in Fig. \ref{fig:nperpsquaredvsdensity} for a deuterium plasma in a \SI{1.4}{T} magnetic field with $n_\parallel$ at the antenna. It is clear that the fast wave can only propagate at densities greater than roughly \SI{2e18}{m^{-3}}, whereas the slow wave can propagate at densities as low as \SI{3e16}{m^{-3}} and is limited to densities below the lower-hybrid resonance. This is one reason it was important to minimize the vacuum gap between antenna and plasma, to ensure that the density at the plasma edge was sufficiently high for fast wave propagation.
\par
An accurate determination of the total refractive index $\vec{n} = \vec{k} c/\omega$ where $|\vec{n}| = n = \sqrt{n_\parallel^2 + n_\perp^2}$ is important for calculating the wave-particle resonance condition, which is discussed further in Sec \ref{subsec:resonance_condition}. To a reasonable degree of accuracy, conservation of toroidal mode number allows $n_\parallel$ to be estimated at any major radius using its value at the antenna, $n_{\parallel, antenna}$, and the major-radial position of the wave at a given point. In other words, because of axisymmetry, the number of wavelengths along the toroidal direction is conserved as the wave propagates. This means that as the waves propagate inwards from the antenna on the low-field side to the tokamak interior, i.e. from larger to smaller major radius, the toroidal wavelength will decrease. This causes an up-shift in $k_\parallel$. Taking the antenna to be at a major radius of \SI{2.27}{m} and the magnetic axis to be at \SI{1.67}{m}, $k_\parallel$ increases by a factor of $2.27/1.67 \approx 1.36$.
\par
\begin{figure}
    \centering
    \includegraphics[width=\columnwidth]{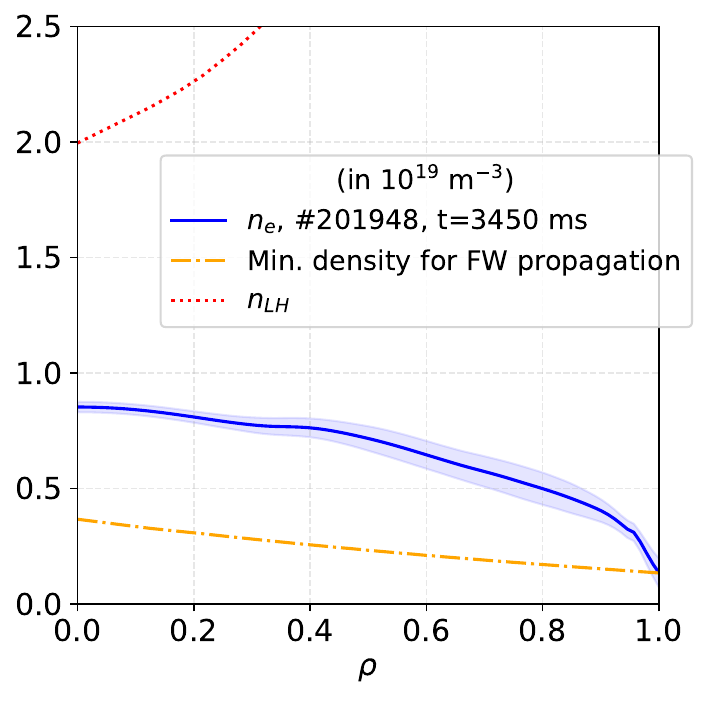}
    \caption{The minimum density for fast wave propagation as a function of normalized minor radius $\rho$ for DIII-D shot 201948 at \SI{3.45}{s} (solid orange line). The measured density profile from Thomson scattering and interferometry is also shown (blue line). The density at which the lower hybrid resonance occurs, $n_{LH}$, is also shown (dashed red line in the top left of the figure).}
    \label{fig:density vs rho}
\end{figure}

To investigate further the minimum density the fast-wave can propagate, we assume $B(R) \approx B_0 R_0/R$ and $n_\parallel(R) \approx n_{\parallel, antenna} R_{antenna}/R$, where the subscript zero denotes values on the magnetic axis ($R_0 \sim \SI{1.7}{m}$ and $B_0 \sim \SI{1.4}{T}$). Since both vary across a poloidal cross-section, we use the EFIT equilibrium to map R to normalized minor radius $\rho$ and then plot in Fig \ref{fig:density vs rho} the minimum density the fast-wave (FW) can propagate at. For the FW minimum density calculation, at each value of $\rho$ the cold plasma dispersion is solved with the appropriate values of $B$ and $n_\parallel$. Fig \ref{fig:density vs rho} shows that the density of QRE shots was sufficiently high throughout the profile to sustain FW propagation. The density profile shown is a smoothed profile using both Thomson scattering and interferometer data. The density at which the lower hybrid resonance occurs, $n_{LH}$ for a given magnetic field and wave frequency (f = \SI{476}{MHz}) is also shown. It is the maximum density the slow wave can propagate and can be seen to be much higher than the experimentally measured density profiles. Hence, incidentally launched or mode-converted slow waves may also be present throughout the entire plasma volume in the investigated QRE plasmas.
\par


\par
\begin{figure}[htbp]
    \centering
    \includegraphics[width=\columnwidth]{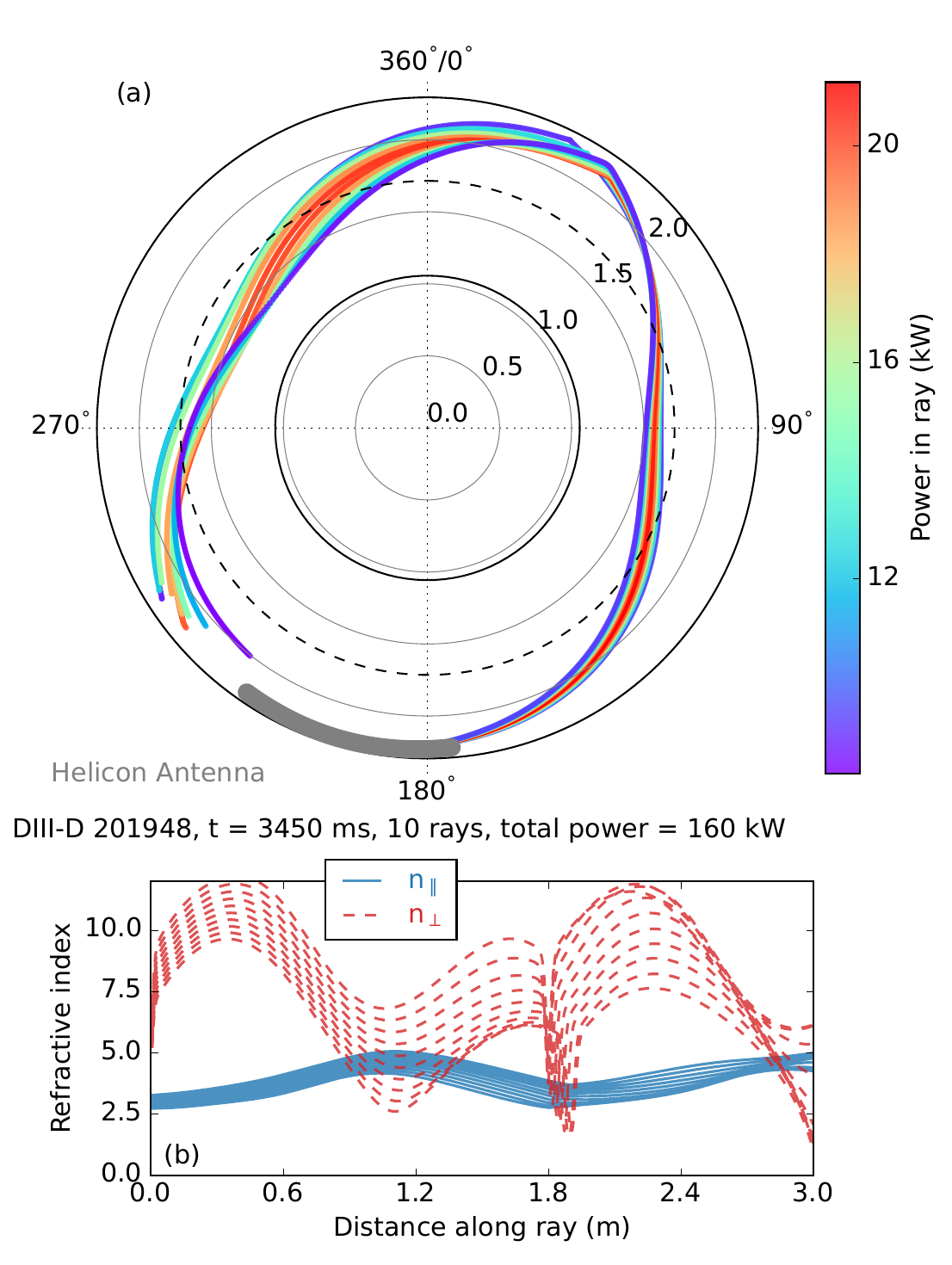}
    \caption{(a) A top-down toroidal projection of the ray trajectories as calculated by GENRAY raytracing simulations for DIII-D shot 201948. The location of the antenna is shown schematically in grey. (b) The evolution of $n_\parallel$ and $n_\perp$ along the trajectory. The rays are simulated until they reflect twice off the last closed flux surface or travel a pre-set distance poloidally, whichever comes first.}
    \label{fig:nparandnperp_vs_distancealongray}
\end{figure}
\begin{figure}
    \centering
    \includegraphics[width=\columnwidth]{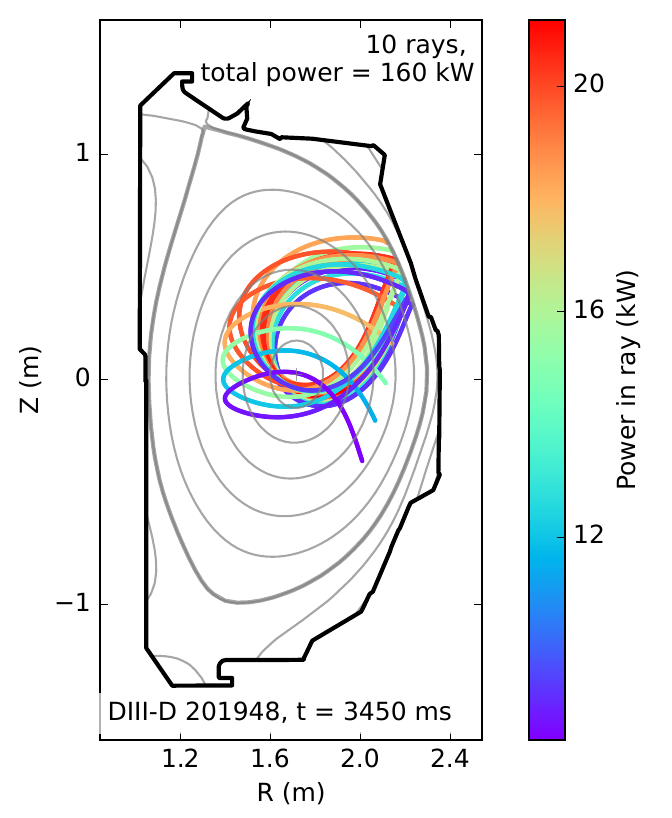}
    \caption{Poloidal projections of 10 ray trajectories simulated in GENRAY. The rays are initialised with slightly different $n_\parallel$ values, centred around $n_\parallel=3$. The waves initially travel downwards from the antenna and spiral into and back out from the core. One limitation of GENRAY is that it handles reflections like light off of a mirror, in that the wave vector gets flipped and then propagation continues as before.}
    \label{fig:poloidal_trajectory_genray}
\end{figure}

Determining $n_\perp$ can be done by analytically solving the cold-plasma dispersion relation, but is most accurately done by raytracing simulations using the GENRAY code \cite{smirnovgenrayraytracing2001}. We show the results of GENRAY simulations for the fast wave in Figs \ref{fig:nparandnperp_vs_distancealongray} and \ref{fig:poloidal_trajectory_genray}. 10 rays are used in the simulation, where a total input power of $\SI{160}{kW}$ is split between the 10 rays according to an initial gaussian power spectrum. The exact powers of the rays are shown in Fig \ref{fig:npara_power_spectrum} in Appendix \ref{app_sec:npara_power_spectrum}.
The evolution of $n_\parallel$ and $n_\perp$ from GENRAY for shot 201948 is shown in panel (b) of Fig. \ref{fig:nparandnperp_vs_distancealongray}, and the toroidal trajectory of the rays and power deposition are shown in panel (a). The trajectory of the rays in a poloidal projection is shown in Fig \ref{fig:poloidal_trajectory_genray}. The fast waves certainly make it into the core where most REs will be with a significant fraction of their initial power.

\subsection{Relativistic Wave-particle resonance}
\label{subsec:resonance_condition}

Wave--particle interactions govern how externally launched waves can resonate with
relativistic electrons in tokamaks. For an electron of velocity $\mathbf{v}$
interacting with a wave of frequency $\omega$ and wave-vector $\mathbf{k}$, the
relativistic cyclotron resonance condition is
\begin{equation}
\omega - \mathbf{k}\cdot\mathbf{v} = \frac{n\,\Omega_{ce}}{\gamma},
\label{eq:res_general}
\end{equation}
where $n$ is the harmonic number,
$\Omega_{ce}=|eB/m_e|$ is the (positive) electron cyclotron frequency,
and $\gamma = 1/\sqrt{1 - v^2/c^2}$ is the Lorentz factor.
\par
As shown in App \ref{app:pitch-angle-derivation}, Eq. \ref{eq:res_general} can be used to calculate the pitch-angle cosine $\xi = v_{\parallel}/v$ of a RE of kinetic energy $E_k$ with relativistic momentum $p(E_k)$:
\begin{equation}
\xi(E_k) =
\frac{m_e\big[\gamma(E_k)\,\omega - n\,\Omega_{ce}\big]}
{k_{\parallel}\,p(E_k)}.
\end{equation}
\par
\begin{figure}[htbp]
    \centering
    \includegraphics[width=\columnwidth]{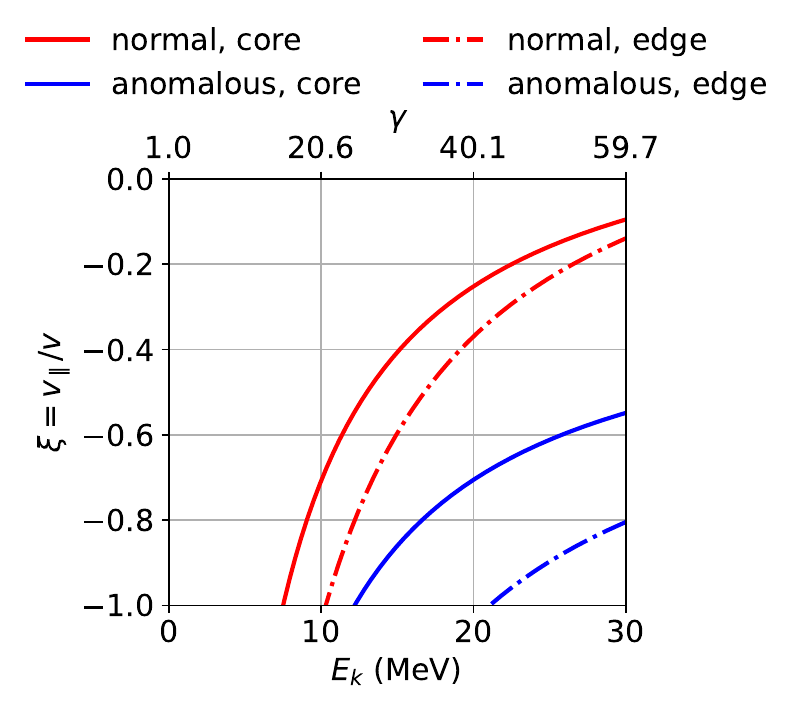}
    \caption{A comparison of the normal (n=1) and anomalous (n=-1) Doppler-shifted cyclotron resonances for \SI{476}{MHz} helicon waves in a \SI{1.44}{T} background magnetic field. For clarity, the anomalous resonance lines are shown for $\mathbf{k} \cdot \mathbf{v} > 0$  whereas the normal resonance lines are for $\mathbf{k} \cdot \mathbf{v} < 0$ (so the y-axis is technically $\xi = - |v_{\parallel}|/v$). The core and edge curves correspond to $|k_\parallel| = \SI{44}{m^{-1}}$ and $|k_\parallel| = \SI{30}{m^{-1}}$ respectively, which are estimated from GENRAY raytracing simulations. Note that for all resonance curves, the core magnetic field of \SI{1.44}{T} is used, so that the effect of different $k_{\parallel}$ on the resonance with REs in the core is clear.
    }
    \label{fig:xi_vs_E_948}
\end{figure}

This expression provides a direct mapping between the kinetic energy of an
electron and the pitch angle required to satisfy the cyclotron resonance for a
wave with given $\omega,k_{\parallel}$ and $n$. Using the estimated $k_{\parallel}$ values from GENRAY simulations for both core and edge values, Fig.\,\ref{fig:xi_vs_E_948} shows the resonant pitch angle as a function of kinetic energy for the normal ($n=1$) and anomalous ($n=-1$) Doppler-shifted cyclotron resonances for helicon waves at \SI{476}{MHz} in a \SI{1.44}{T} magnetic field. This figure also conveys why targeting the normal resonance is more desirable than the anomalous resonance for RE scattering, since it resonates with lower energy electrons and so can `clip' the RE vortex at lower energies \cite{guocontrolrunawayelectron2018}. It also illustrates that the up-shift in $k_{\parallel}$ as the wave propagates inwards from the antenna \textit{improves} the wave's ability to scatter REs of lower energy. 
\par
There is a neat way to see why the $n = 1$ resonance can be satisfied at lower energies than the $n = -1$ resonance: Eq \ref{eq:res_general} can be rearranged to be $|k_\parallel v_\parallel| = |\omega - n \omega_{ce}/\gamma|$. For fixed $\omega$ and $\omega_{ce}$, the right-hand side ($|\omega - n \omega_{ce}/\gamma|$) will always be smaller for $n = 1$ than $n = -1$, independent of the size of $\gamma$, so for fixed $|k_\parallel|$, this will make $|v_\parallel|$ smaller for $n = 1$ than $n = -1$, independent of the size of $v_\perp$, allowing that resonance to be satisfied at lower energies.
\par
In general, for a given $\mathbf{k}$, either the normal ($n=1$) or anomalous ($n=-1$) will be dominant in the region of phase space where REs exist. As shown in Fig 1 (b) of Ref \cite{choudhuryFirstDemonstrationResonant2026}, most REs have $\xi \approx -1$. Since launching waves of different direction flips the sign of $k_\parallel$ in Eq \ref{eq:res_general}, the sign of $k_\parallel$, for a fixed configuration of $I_p$ and $B_T$, determines the dominant resonance. Crucial for choosing the sign of $k_\parallel$ is recognizing that the $n=1$ resonance can only be satisfied if $k_\parallel \cdot v_\parallel < 0$. If instead $k_\parallel \cdot v_\parallel > 0$, then either the $n=0$ or $n=-1$ resonances can be satisfied. Since the $n=0$ case is the Landau damping (also called Cherenkov resonance) case, which would occur at thermal velocities, there are no interesting RE dynamics there. Therefore, to summarize, fixing $I_p$, $B_T$, and $k_\parallel$ completely determines the primary RE-wave resonance.
\par
More detailed quasi-linear simulations of the wave-particle interaction in phase space which will take into account both the spectrum in $\vec{k}$ and the RE distribution function are possible \cite{guocontrolrunawayelectron2018,choudhuryFirstDemonstrationResonant2026}, though are limited by assuming a certain wave energy density. It's unfortunately not yet possible to accurately estimate the wave energy density in these experiments, since the fraction of coupled power that makes it into the plasma is unknown, as is the volume over which the wave energy is deposited. Those quasi-linear simulations generally find that above some threshold wave energy density, the diffusion in RE phase space does not change drastically. There are ongoing efforts to use the Doppler Back-Scattering device \cite{chowdhuryNovelDopplerBackscattering2023} on DIII-D to measure the helicon electric field directly \cite{Chowdhury2026HeliconProfiles}, which if successful could be used to inform a reasonable wave energy density for simulations of these experiments.

\subsection{Experimental Configurations}
Throughout this paper we take plasma current direction to be fixed in the standard direction. Since the toroidal field direction can either be in the ideal or non-ideal configuration and there are two possible directions to launch the wave, either parallel or anti-parallel to the plasma current (or, equivalently, either anti-parallel or parallel to the bulk RE velocity, $\mathbf{v_{RE}}$), there are four possible experimental configurations shown in Table \ref{tab:four_cases}. One configuration, that of the non-ideal case with $\mathbf{k} \cdot \mathbf{v_{RE}} < 0$ which primarily targets the anomalous resonance, was not explored in these experiments owing to restrictions in runtime and antenna conditioning.
\begin{table}[h]
\centering
\begin{tabular}{c|c|c}
 & Non-ideal & Ideal \\
\hline
$\mathbf{k} \cdot \mathbf{v_{RE}} > 0$ & (Anomalous)& Anomalous \\        
\hline
$\mathbf{k} \cdot \mathbf{v_{RE}} < 0$ & Normal & Normal \\        
\end{tabular}       
\caption{The four possible experimental configurations explored and the primary wave-particle resonance targeted in each case. The bracketed top-left quadrant was not experimentally explored, while the remaining configurations were. The plasma current direction is fixed to be anticlockwise as viewed from above. A schematic of the directions of $k$, $B_T$ and $I_p$ in the ideal configuration is shown in Fig \ref{fig:topdownschematic}.}
\label{tab:four_cases}
\end{table}

\subsection{Main RE diagnostics}
REs emit several distinct types of radiation which can be measured passively on DIII-D and be used to infer properties of the RE population. These types of emission are Bremsstrahlung, cyclotron, and synchrotron emission. 
\par
As REs move through the plasma they undergo small-angle Coulomb collisions with background plasma ions and release Bremsstrahlung radiation, typically in the hard X-ray (HXR) range of energies (roughly \SIrange{0.1}{10}{MeV}). In disruption experiments, when REs strike plasma-facing components, they can also emit bremsstrahlung radiation, but this is not the case in QRE experiments since the REs are well confined. The bremsstrahlung radiation can be measured using plastic scintillator detectors. These detectors were primarily developed to measure \SI{2.5}{MeV} neutrons released from deuterium fusion, but can also measure HXR radiation from REs \cite{heidbrinkneutronfluctuationmeasurements1986,heidbrinkmeasurementsneutronsource1997}. The HXR scintillator signal is equal to $\int\int dp d\xi W(p,\xi)F(p,\xi)$, where $W$ is the scintillator-specific weight function and $F$ is the RE distribution function written using (p,$\xi$) as the chosen velocity-space variables. The weight function $W$ is not known exactly for the scintillators used, though there are efforts to determine it, and so comparisons between plastic scintillators, which have different energy sensitivities, is not possible. Even if $W$ were known, the several plastic scintillators on DIII-D are not all co-located and have different lines of sight. Another experimental challenge encountered is the low signal-to-noise ratio for the detectors. This arose because the ideal configuration of reversed $ B_T$ and standard $ I_p$ is not the typical QRE scenario (the non-ideal configuration is). This meant new error-field correction had to be done and since experimental runtime was limited, we were unable to achieve the error-field configuration which is available in the non-ideal configuration. All of the above means the HXR scintillator data is only used to infer qualitative trends in the energy-weighted RE population.
\par
(It is worth repeating that by `ideal' configuration we mean one of ideal antenna alignment with the edge magnetic field, which can either of 1) reversed $ B_T$ and standard $ I_p$ or 2) standard $ B_T$ and reversed $ I_p$.)
\par
DIII-D does, however, have an array of collimated HXR detectors, known as the Gamma-Ray Imager (GRI) which is able to measure the energy spectrum of bremsstrahlung radiation from REs \cite{lvovskiyupgradesgammaray2022,lvovskiyinvestigationrunawayelectron2017,pacegammarayimager2016,cooperapplyingnewgamma2016,paz-soldanresolvingrunawayelectron2018}. The energy spectrum of the gamma rays (i.e. HXRs) is inferred by analysing the heights of measured pulses over a certain time window. A large enough time window is necessary to get good statistics. In Ref \cite{choudhuryFirstDemonstrationResonant2026}, a large time window of $\sim \SI{400}{ms}$ was conservatively used to ensure good statistics. Having confirmed shorter pulses are still accurate, however, the GRI results shown in this paper use much smaller time windows. The GRI spectrum can be written as $\langle \int d\xi W(p,\xi)F(p,\xi) \rangle$, where the triangular brackets denote a time average, and $W(p,\xi)$ is the GRI-specific weight function.
\par
To clarify, both the HXR scintillators and the GRI measure bremsstrahlung from the REs, but only the GRI can give energy-resolved spectra vs time. We therefore use the GRI to make quantitative statements about the energy distribution of the REs, and the scintillators to make qualitative statements about the growth, plateau, or decay of the RE population as a whole. 
\par
REs also emit cyclotron radiation as they spiral around magnetic field lines. This radiation is typically in the microwave range of frequencies for DIII-D REs and can be measured using heterodyne radiometers \cite{austinelectroncyclotronemission2003}. There are 40 chords to the radiometer which cover different radial locations of the plasma. In this paper, unless otherwise stated, we use the innermost chord, which is the one with the smallest major radius.
\par
\begin{figure}
    \centering
    \includegraphics[width=\columnwidth]{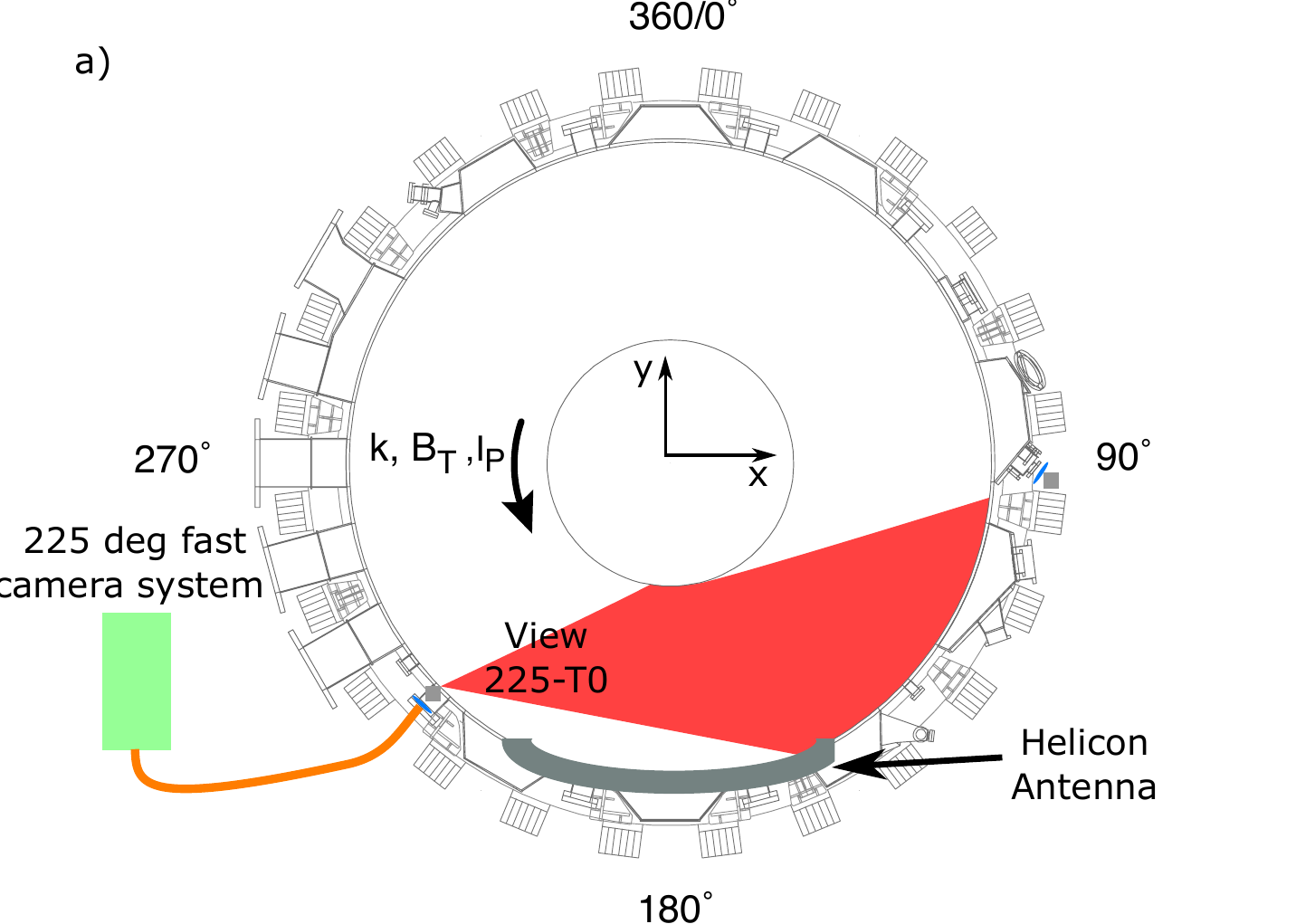}
    \caption{Top-down schematic view of the DIII-D tokamak. The arrows for $k, B_T$ and $I_p$ show the ideal configuration with $\vec{k} \cdot \vec{v_{RE}} < 0$ which primarily targets the normal Doppler resonance. The view of the fast camera is also shown as well as the position of the helicon antenna.}
    \label{fig:topdownschematic}
\end{figure}

Finally, REs emit infrared synchrotron radiation in a narrow beam in the parallel direction as they travel at relativistic speeds \cite{hoppeinterpretationrunawayelectron2018,stahlSynchrotronRadiationRunaway2013}. The fast camera diagnostic on DIII-D is able to capture images of this synchrotron radiation every \SI{5}{ms} using a filter that only passes light of wavelengths between \SI{700}{nm} and \SI{800}{nm} \cite{marinifastcamerafastcam2023}. Theoretical work in Ref \cite{stahlSynchrotronRadiationRunaway2013} has predicted wave particle interactions to increase the measurable infrared synchrotron emission, and on DIII-D QRE plasmas, RE-driven kinetic instabilities have been observed to scatter the REs that excited them, resulting in oscillations in the synchrotron emission that resemble a predator-prey cycle \cite{choudhurydetailedcharacterizationrunaway2025}. The fast camera is therefore a very useful diagnostic to observe changes in the RE population that result from wave-particle interactions. The viewing geometry of the fast camera is shown in Fig \ref{fig:topdownschematic} and the arrows for the wavevector $\vec{k}$, $ B_T$ and plasma current $ I_p$ show the ideal configuration with $\vec{k} \cdot \vec{v_{RE}} < 0$ which primarily targets the normal Doppler resonance.

\section{Results}


Before discussing the main experimental findings, Sec \ref{sec:effects_of_helicon_deltavloop} describes the effect of the antenna's operation on density and loop voltage. 
Then Sec \ref{sec:further_details_of_normal_in_ideal} adds further details of the effectiveness of targeting the normal resonance in the ideal configuration; Sec \ref{sec:effect_of_wave_direction} discusses the effect of changing the direction of the launched helicon waves; and Sec \ref{sec:effectiveness_of_normal_in_nonideal_case} describes the effectiveness of targeting the normal resonance in the non-ideal experimental configuration. 

\subsection{Effects of helicon antenna on density and loop voltage}
\label{sec:effects_of_helicon_deltavloop}
\begin{figure*}
    \centering
    \includegraphics[width=\textwidth]{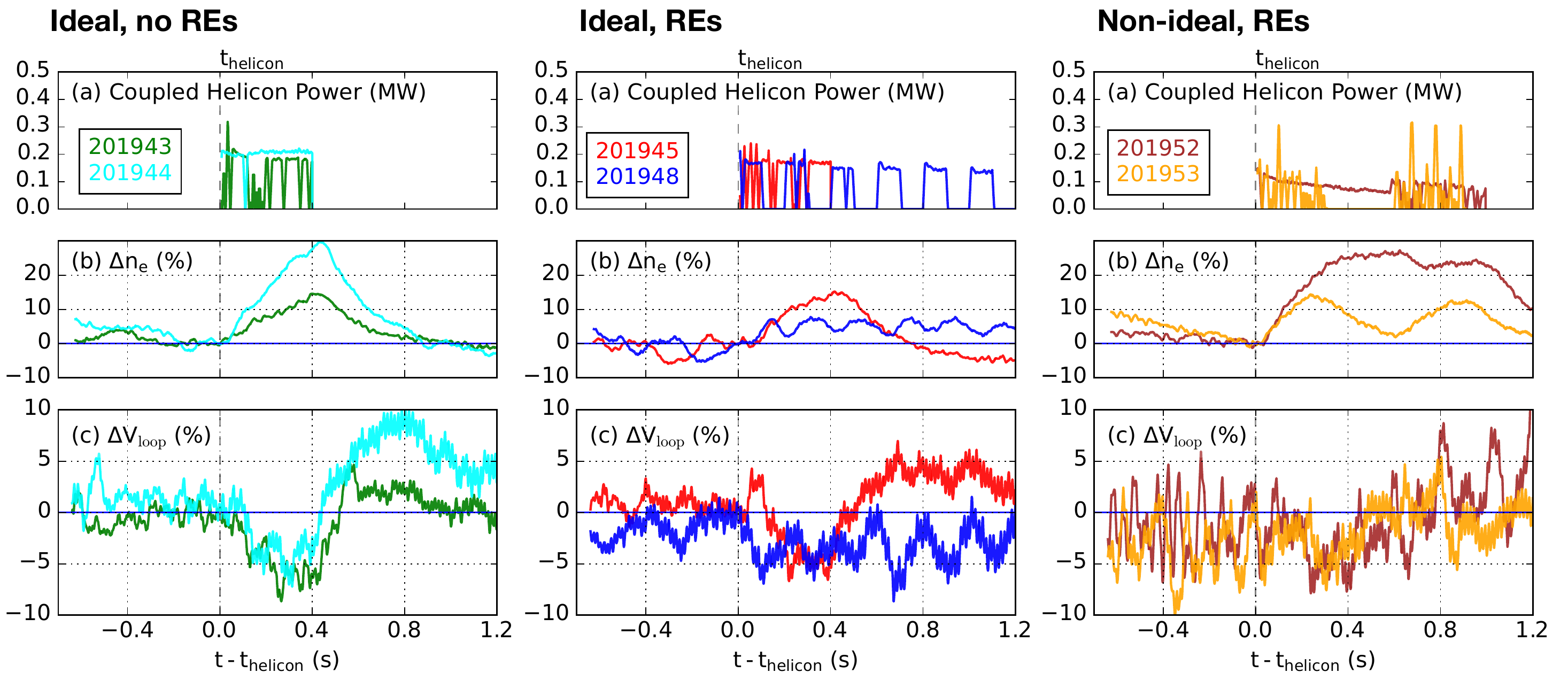}
    \caption{The effect of helicon power on three different cases: ideal plasmas with no REs; ideal plasmas with REs; and non-ideal plasmas with REs. Panel (a) in each case shows the coupled helicon power. Both (b) and (c) show the percentage changes of the signals normalised to their average values in the \SI{30}{ms} preceding the start of helicon power. (b) shows the percentage changes in density, as measured using the CO2 interferometer with a vertical line-of-sight near the plasma core. (c) shows the percentage changes in $V_{loop}$.} 
    \label{fig:deltavloop}
\end{figure*}

The effects of helicon operation on density and loop voltage, as compared to their values just before helicon power is turned on, are shown in Fig \ref{fig:deltavloop}. Operation of the antenna was empirically observed to increase the electron density by up to as much as 20\% in a range of shots and circumstances, including both ideal and non-ideal shots, and shots with and without REs. In addition, there also seem to be smaller, up to 10\%, changes to the loop voltage, but these do not have a clear pattern in all cases except for the ideal case with no REs; in that case, the loop voltage decreases by 5\% during helicon power, and then increases to 5\% higher than its original level once the power is turned off. Since $E/E_c \sim V_{loop}/n_e$, the density increases will decrease $E/E_{c}$ to first order by the percentage change in density. 
\par
The exact cause of the increases in density cannot be definitively stated. The authors suspect it is neither from impurities from the helicon antenna itself, nor from ionization of neutrals, since QRE plasmas generally are fully ionized. One possibility is that there is desorption from the antenna's surface when helicon power is on. Whatever the cause of the measured density increase, which has also been observed during helicon operation in non-QRE plasmas with no REs, it will certainly affect the HXR scintillator signal, since there are more background ions for the REs to scatter off and release bremsstrahlung.
\par
Throughout the results section, it is worth distinguishing the short-time scale effects on the measured signals during each pulse of helicon operation, if it is indeed pulsed, and the longer, order one second, behaviour of the time signals. For example, in Fig 2 of Ref \cite{choudhuryFirstDemonstrationResonant2026}, there is a short-time scale initial increase in the HXR scintillator signal. Unless otherwise specified, we will focus on the longer-time scale results.

\subsection{Further details of the Normal Resonance in the Ideal-Bt Configuration}
\label{sec:further_details_of_normal_in_ideal}
\par
\begin{figure}
    \centering
    \includegraphics[width=\columnwidth]{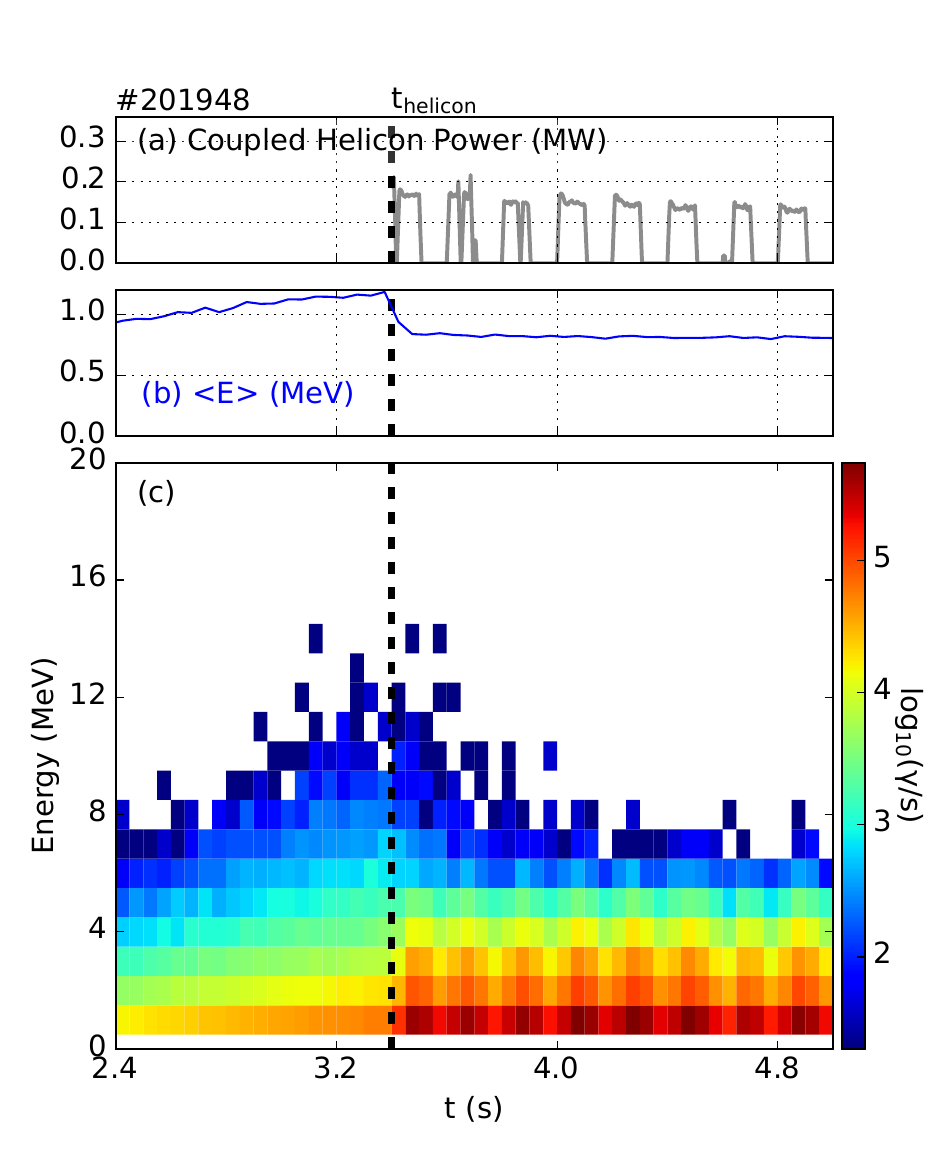}
    \caption{Detailed energy spectra vs time for shot 201948. (a) Coupled helicon power. (b) and (c) are from the Gamma Ray-Imager diagnostic: (b) The arithmetic average of the HXR energies; (c) Energy spectra of the incident HXRs over \SI{50}{ms} wide windows. Note the scale of the color-map is $\log_{10}$ and normalized to give the number of HXRs per second. The white area in (c) are those parts of the spectra for which no HXRs were measured. The energy bins are \SI{1}{MeV} wide and start at \SI{0.5}{MeV}. 
    }
    \label{fig:201948energyvstimecontour}
\end{figure}

 Fig \ref{fig:201948energyvstimecontour} shows the effect helicon power, panel (a), had on a representative QRE plasma. The average energy of the REs, panel (b), and the energy spectrum of REs, panel (c), as measured by the GRI are also shown. The GRI measurements here used a \SI{50}{ms} time window. After the antenna is turned on, there is a clear drop in REs with energies above \SI{8}{MeV}, and an increase in REs with energies below \SI{4}{MeV}. Without any helicon power, the RE energy spectra of these QRE plasmas with high $E/E_c$ shift to higher energies as time progresses (see Sec \ref{sec:QREscenario} for a discussion of the QRE growth phase). 
\par
It is interesting that low-energy HXR counts fluctuate in number, though the high-energy HXR counts do not. This could be because there are several orders of magnitude more low-energy HXRs than high-energy HXRs, so small changes in the low-energy RE population are more easily observed, whereas following helicon there is no measurable population of REs above \SI{8}{MeV} left. The average HXR energy shown in panel (b) of Fig. \ref{fig:201948energyvstimecontour} also clearly drops following helicon turn-on, confirming the trend seen in panel (c).
\par
\begin{figure*}
    \centering
    \includegraphics[width=\textwidth]{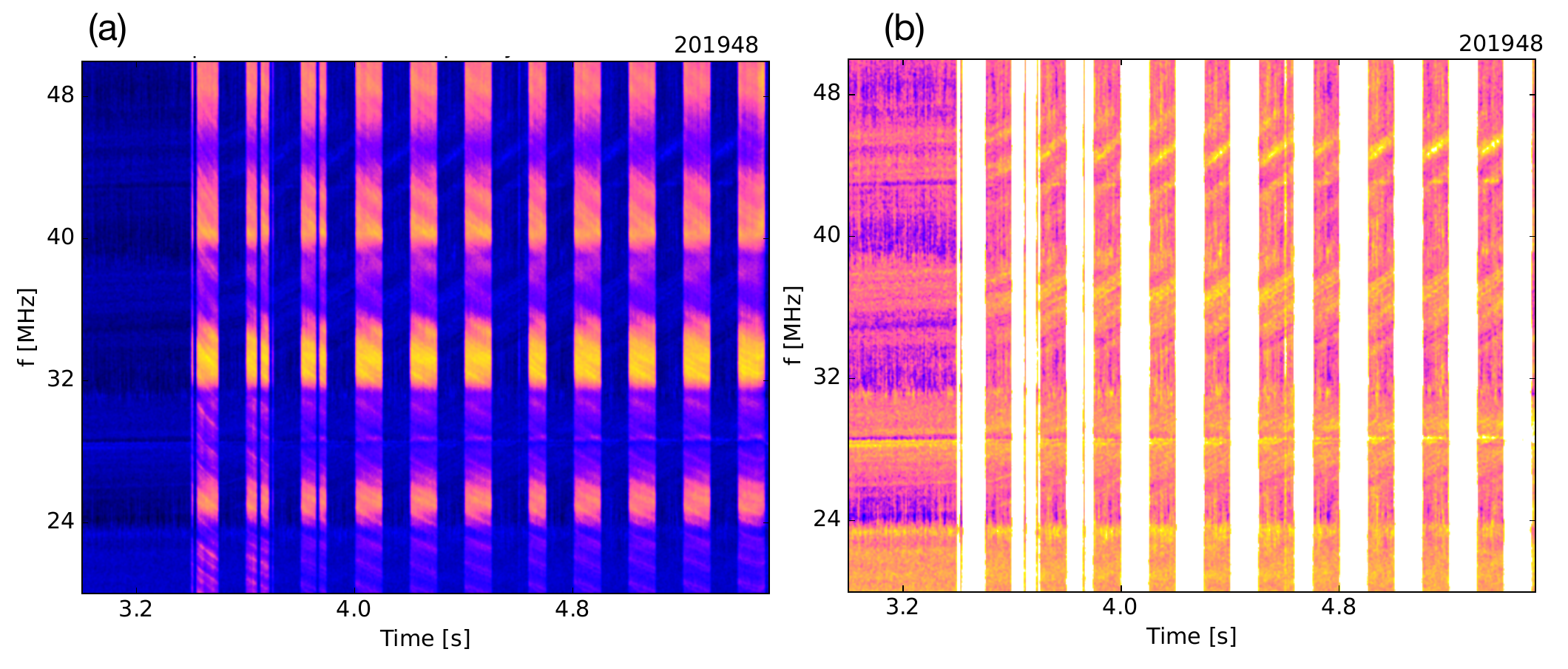}
    \caption{Fast Fourier transform spectrograms of the fast magnetic loops on DIII-D. This particular loop is on the outer wall and measures in the frequency range \SIrange{0}{100}{MHz}, though only \SIrange{20}{60}{MHz} is shown here. The loop is not absolutely calibrated so both color maps are in arbitrary units. Higher power fluctuations are shown in panel (a) which appear saturated in panel (b) which has a colorbar that shows lower power modes. }
    \label{fig:icei_201948}
\end{figure*}

Another interesting observation from shot 201948 is from the so-called ICE diagnostic, that has historically been used to measure instabilities in the ion cyclotron range of frequencies on DIII-D \cite{degrandchampioncyclotronemission,thomecentralioncyclotron2019,heidbrinkdependenceioncyclotronrange2025,heidbrinkdependenceioncyclotronrange2025a,heidbrinkdependenceioncyclotronrange2025b,degrandchampupgradesioncyclotron2021,tangStabilizationAlfvenEigenmodes2021},. The ICE diagnostic consists of several conducting loops with sampling rates of \SI{200}{MHz}, and hence a Nyquist frequency of \SI{100}{MHz}. This diagnostic has been used in the past to detect the parametric decay instability during helicon power \cite{pinskerfirsthighpowerhelicon2024}. Shown in Fig \ref{fig:icei_201948} are two spectrograms from one such loop for shot 201948. They are represent the same data but use a different color scale to highlight higher power fluctuations, in panel (a), and lower power fluctuations, in panel (b).  During the helicon-on times, the loop measures very strong broadband fluctuations over the whole range of frequencies. This is most likely due to the parametric decay instability of the helicon wave into lower frequency waves. In (a), note that there are strong modes separated roughly 8 MHz apart. The deuterium cyclotron frequency $eB/(2\pi m_D)\approx\SI{7.6}{MHz}$ at the plasma edge where $B\approx\SI{1}{T}$ and $m_D$ is the mass of deuterium. This suggests these modes are indeed from the parametric decay instability. In panel (b) it can be seen that during the off periods there are rising tones that seem to increase in frequency over a $\sim\SI{5}{MHz}$ range. They are not observed before the antenna is turned on, nor are they observed outside the frequency range shown on the spectrogram (the fast magnetics have a sampling rate of \SI{200}{MHz}, so frequencies between \SIrange{0}{100}{MHz} can be observed). The exact cause of these modes is not known and remains an object for future study. One possibility for the frequency sweeping is the slight changes in position caused by helicon heating and subsequent changes in the Shafranov shift. The on-periods in panel (b) appear as white vertical bars because of the color-bar chosen. Those features were allowed to appear saturated to bring out the features during the off-periods. 

\par
\begin{figure}
    \centering
    \includegraphics[width=\columnwidth]{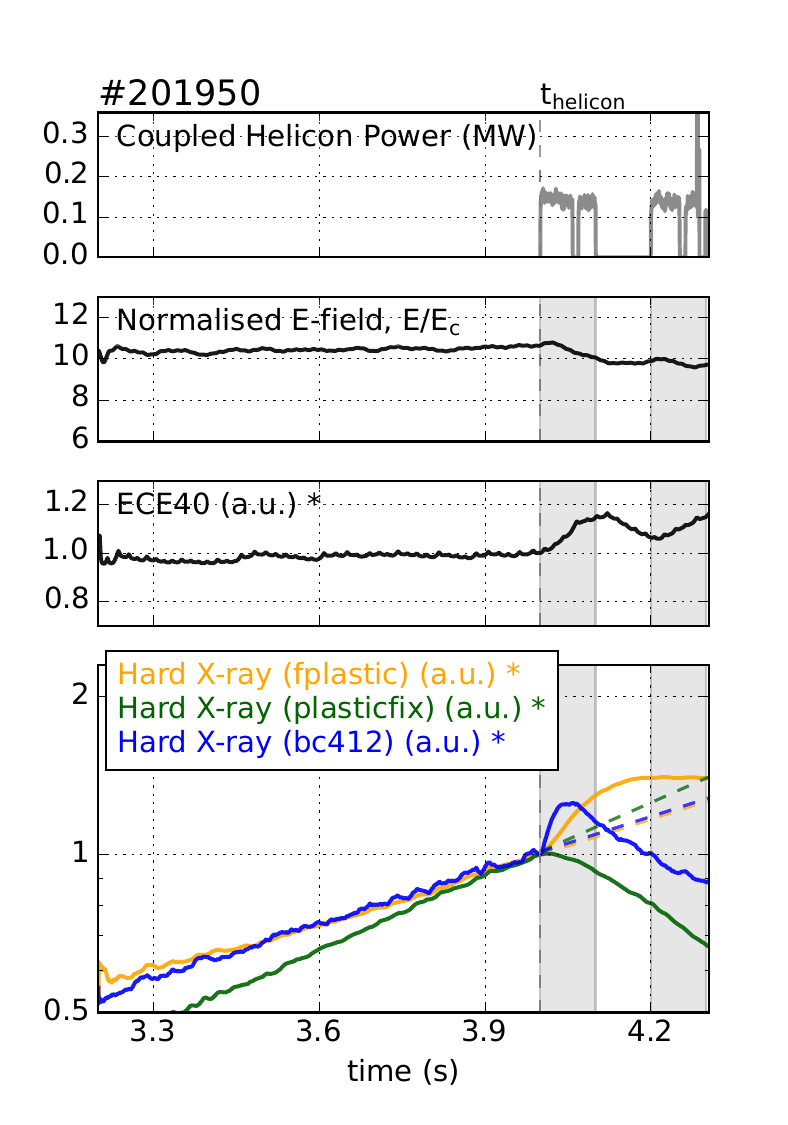}
    \caption{The interesting case of 201950. Another ideal shot that had helicon targeting the normal wave particle resonance. Both the electron cyclotron emission (ECE40), and the scintillator signals are normalised to their values just before helicon power starts.}
    \label{fig:201950_timetraces}
\end{figure}
\begin{figure*}
    \centering
    \includegraphics[width=\textwidth]{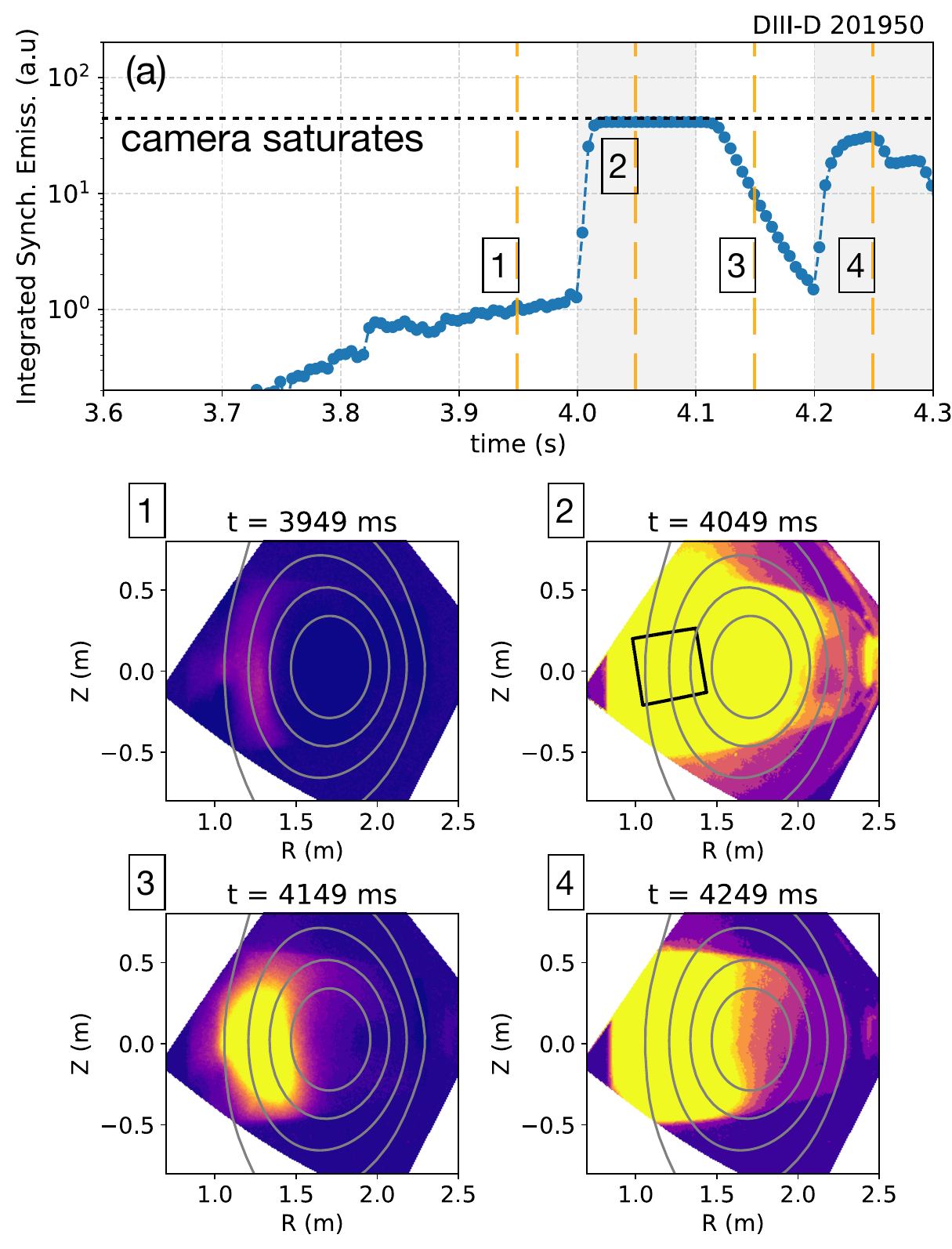}
    \caption{Images 1-4 show RE synchrotron emission before and during the first two helicon pulses of DIII-D shot 201950, with the EFIT magnetic equilibrium shown too. Integrating the pixel counts over the black box shown in image 2 results in a 1D time-trace of the integrated synchrotron emission, shown in the top panel (a). The exact times of the images 1-4 are shown as orange vertical lines in panel (a). The images shown are from the \SI{225}{\degree} camera system whose view is shown in Fig \ref{fig:topdownschematic}.}
    \label{fig:201950_synchrotron_emission}
\end{figure*}

Because DIII-D has a carbon wall, there are shot-shot variations in the density achievable in QRE shots. One shot, 201950, had a particularly low density and hence a high $E/E_{c}$ which over four seconds was able to produce a significant RE population. Key time traces for 201950 are shown in Fig \ref{fig:201950_timetraces}. The bottom-most panel shows three independent scintillator signals, where each signal has been normalised to its value at $\textrm{t}=\SI{4}{s}$ which is the time helicon power turns on. In this shot, the coupled helicon power, at around \SI{150}{kW} is comparable to others, but it is unique in the dramatic \textit{decrease} in the scintillator signals. Unfortunately, the error-field correction could not prevent a locked mode in this shot at $\textrm{t}=\SI{4319}{ms}$, and so only \SI{300}{ms} of helicon effects are observable. 
The scattering of REs was so pronounced that the fast camera measuring synchrotron emission completely saturated during the first pulse of helicon, shown in image 2 of Fig \ref{fig:201950_synchrotron_emission}. Three other images are shown: image 1 shows that there was a faint RE crescent on the high-field side \SI{50}{ms} before the first helicon pulse of the shot. Image 2 is right in the middle of the first pulse. Much of the camera's field of view is completely saturated. The camera takes frames every \SI{5}{ms} in this shot. Panel (a) is a time-trace of the integrated synchrotron emission from within the black box shown in image 2. Panel (a) suggests it takes approximately two frames, or \SI{10}{ms}, into the helicon pulse before the camera fully saturates, and then the synchrotron emission gradually reduces over the subsequent \SI{100}{ms} off period. Image 3 shows the brightened RE crescent during the middle of the off period. Then when the synchrotron emission had approximately returned to its pre-helicon level, the second pulse comes in, and although this time not fully saturating the camera, it causes a very pronounced increase in brightness, as seen in image 4. 
\par
\begin{figure}
    \centering
    \includegraphics[width=\columnwidth]{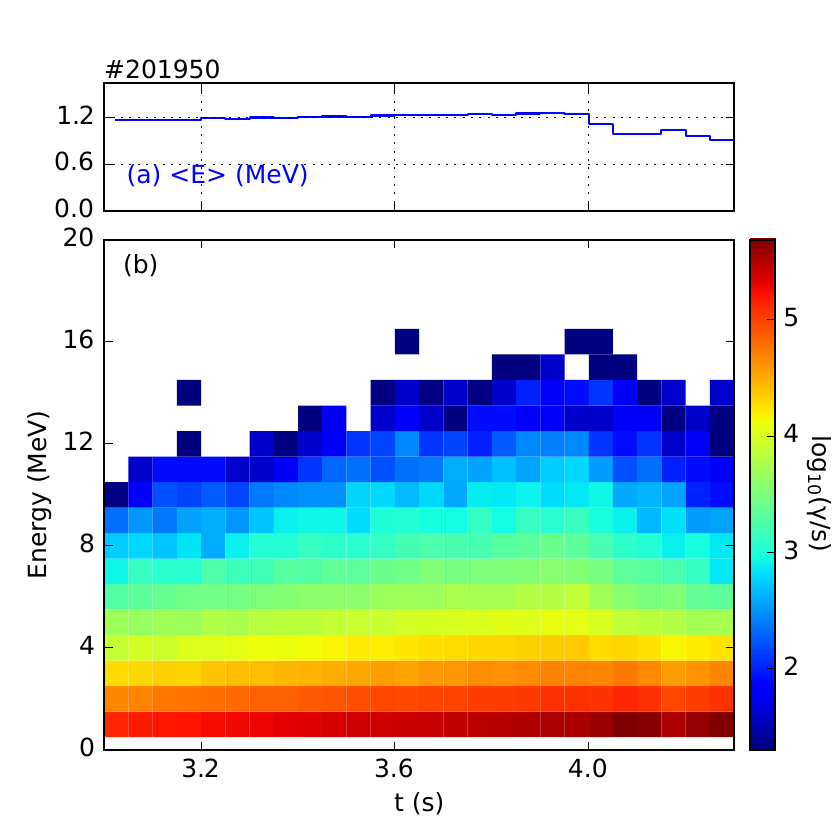}
    \caption{(a) The average energy of RE bremsstrahlung emission as measured by the GRI and (b) the energy resolved spectra for shot 201950.}
    \label{fig:201950energy_contour_plot}
\end{figure}

It was reported in Ref \cite{choudhuryFirstDemonstrationResonant2026} that only the first \SI{100}{ms} pulse of helicon power caused an appreciable increase in synchrotron emission and subsequent ones failed to do so. This was explained as follows: if all the REs of energies greater than the resonance were successfully scattered, the electric field during the \SI{100}{ms} off period (which had a measured value of \SI{0.055}{V/m}) could only accelerate an electron by \SI{1.65}{MeV}. The electric field in shot 201950's case is roughly equivalent at \SI{0.6 \pm 0.05}{V/m}. This suggests that despite significant pitch-angle scattering of REs during the first on pulse, there still remains a significant population of REs around the resonance curve in momentum space for the second pulse to scatter. This suggests that the \SI{160}{kW} or so launched was not sufficient to scatter all the REs and presumably more would have been needed. Since 201950 had the same $B_T$ value, the resonance energy is the same as for 201948 at around \SI{8}{MeV}. The GRI energy spectra for 201950 shown in Fig \ref{fig:201950energy_contour_plot} (b) confirms that indeed there is still a significant RE population with energies greater than \SI{8}{MeV} (c.f. this figure's colorbar scale with that of Fig \ref{fig:201948energyvstimecontour} (c)). The average energy measured by the GRI, panel (a) of Fig \ref{fig:201950energy_contour_plot}, is comparable to that of 201948. This illustrates the importance of not only reducing the average energy of the RE population, which many mitigation schemes focus on, but rather the maximum energy. The scattering in 201948 was very complete and left no significant RE population with energies greater than \SI{8}{MeV} whereas in 201950's case, the scattering was incomplete in that a significant population with energies greater than \SI{8}{MeV} existed beyond the start of helicon power. 
\par


\subsection{The Effect of Wave Direction in the Ideal Configuration}
\label{sec:effect_of_wave_direction}
\begin{figure}
    \centering
    \includegraphics[width=\columnwidth]{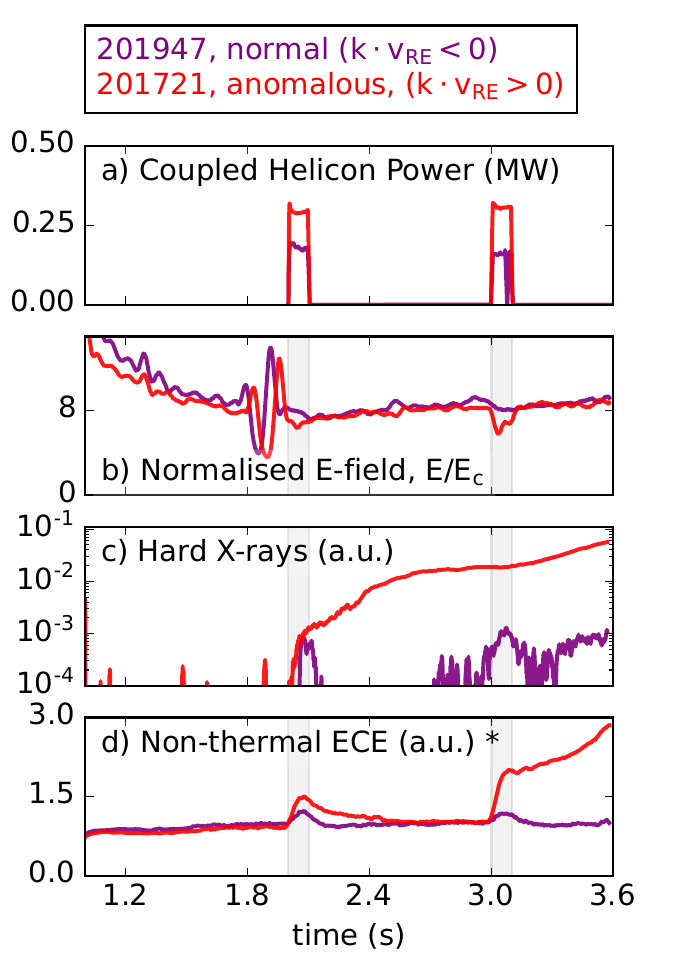}
    \caption{Key time traces comparing two shots with different helicon directions. Despite both shots having similar normalized electric fields, panel (b), the different directions of helicon result in markedly different effects on the HXRs, panel (c). The HXR signals are plotted on a logarithmic scale. The non-thermal ECE signals, panel (d), normalized to a value of 1 at t=\SI{2}{s} for clarity, also differ considerably.}
    \label{fig:normalanomalous}
\end{figure}

Yet another interesting observation concerns two shots, similar in every other respect, that had helicon waves launched in two different toroidal directions. In one case, $\mathbf{k} \cdot \mathbf{v_{RE}} < 0$ and the normal resonance is the dominant resonance in phase space; and the other, with  $\mathbf{k} \cdot \mathbf{v_{RE}} > 0$, strongly favors the anomalous resonance. Shown in Fig \ref{fig:normalanomalous} are the effects that two \SI{100}{ms} helicon pulses, launched at t=\SI{2}{s} and t=\SI{3}{s} had on both shots. We begin by discussing the long-time scale behaviour, i.e., the behaviour over the subsequent second or so from each pulse, and not the diagnostic response during each pulse. The ideal configuration shot, 201947, targeting the normal resonance, showed similar behavior to shot 201948, discussed earlier in this paper and in Ref \cite{choudhuryFirstDemonstrationResonant2026}. The shot favoring the anomalous resonance, however, had markedly different behavior: the HXR signal strongly \textit{increased} and thus suggests the RE population was \textit{increased}, despite both shots having the same normalized electric field. Also in the $\mathbf{k} \cdot \mathbf{v_{RE}} > 0$ shot, the non-thermal ECE approximately \textit{doubles} during the second pulse of helicon, suggesting either a large increase in low-energy REs or a large collective increase in their perpendicular velocities. (The oscillations in normalized electric field prior to the first helicon pulse in both shots are the result of crossover, which is when the primary transformer circuit switches current direction).
\par
There was no GRI data, unfortunately, for shot 201721, nor was the synchrotron emission in either bright enough to detect clear crescents. Though the authors speculate the waves in shot 201721 generate a supra-thermal electron tail which accelerates to relativistic energies in the large normalized electric field, the exact cause of this difference will be investigated in future work and provides opportunities for theoretical work and model validation.  

\subsection{The effectiveness of normal-resonance non-ideal Bt configuration }
\label{sec:effectiveness_of_normal_in_nonideal_case}

Prior to the experiment, it was believed that in the non-ideal configuration, there would be insufficient power directly launched into the fast wave to cause significant RE scattering via the normal Doppler resonance. This was because in the non-ideal configuration, the antenna alignment with the edge magnetic field is poor so the Faraday screen elements are less able to short the longitudinal electric field of the slow wave, see Fig \ref{fig:fieldalignment}. Another factor that suggested weak resonance in the non-ideal configuration is that the vacuum gap between antenna and plasma was larger in the non-ideal configuration shots conducted. As mentioned earlier ideal shots had gaps of $\leq \SI{5}{cm}$ whereas the non-ideal shots that will be shown had gaps of nearly \SI{9}{cm}. This is a larger gap than is typical for DIII-D helicon operation and would further reduce the coupled fast-wave power into the plasma and hence unfavorably alter the balance between fast and slow wave power.
\par
\begin{figure}
    \centering
    \includegraphics[width=\columnwidth]{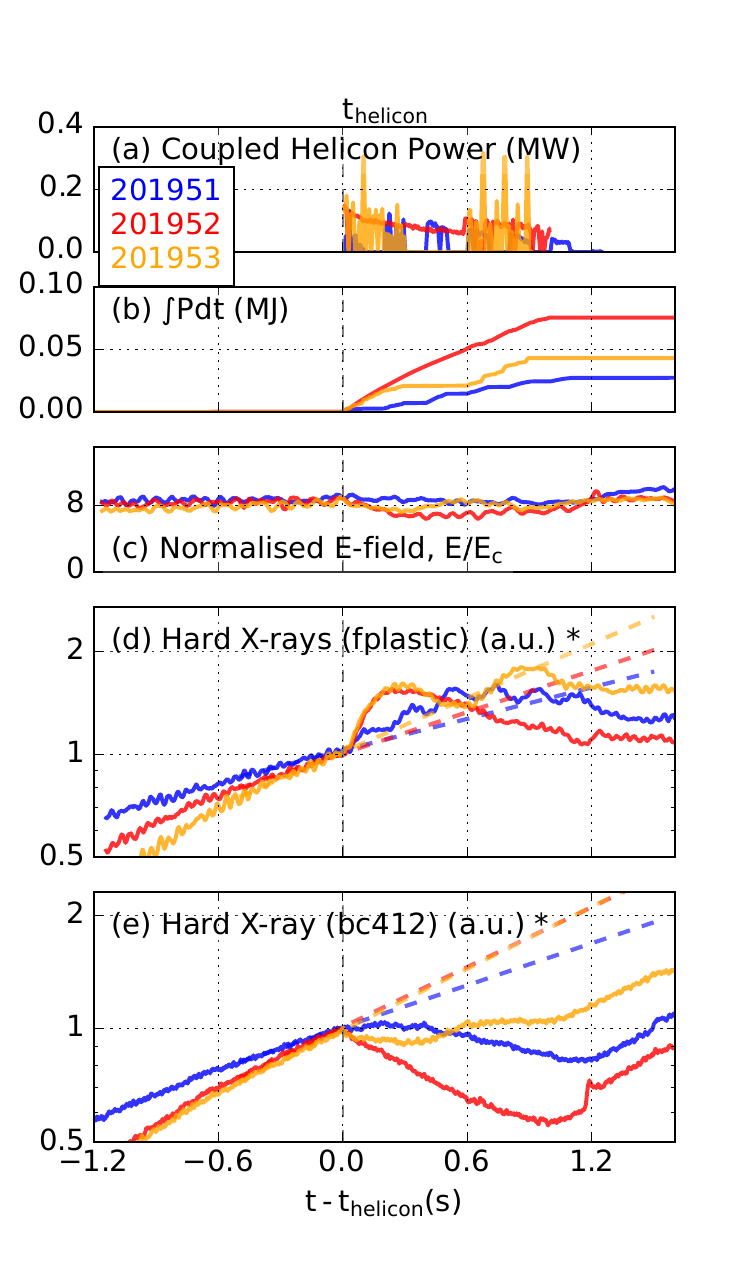}
    \caption{
        Time traces for three non-ideal configuration shots with different helicon waveforms shown in (a). (b) is the time integral of (a) and so shows the cumulative energy injected by the helicon antenna. The normalized electric field, panel (c), is similar in all three shots. In panels (d) and (e), we show the HXR signals from two plastic scintillators at different locations. The HXR signals, all normalized to 1 at the time the helicon power turns on, panel (d), show clear decreases following helicon turn-on. The dashed lines in (d) and (e) show extrapolations of the pre-helicon growth rates for the HXR signals.
    }
    \label{fig:stdBttimetraces}
\end{figure}

Despite both these shortcomings, it has been found that using the helicon antenna in the non-ideal configuration is still very effective at scattering REs. Shown in Fig \ref{fig:stdBttimetraces} are key time traces from DIII-D shots 201951, 201952, and 201953, which were all non-ideal experimental configuration QRE shots which had different helicon waveforms launched. The waves were all launched in the same direction, such that $\mathbf{k} \cdot \mathbf{v_{RE}} < 0$ and the normal resonance is dominant in phase space. The behaviour of the non-ideal configurations is very similar to the ideal configuration shots, and strongly suggests fast-wave scattering. The drop in $E/E_c$ in panel (c) of Fig \ref{fig:stdBttimetraces} can be attributed to increases in density from helicon operation, as discussed in Fig \ref{fig:deltavloop}. The two scintillator signals in (d) and (e) have at first different short timescale behaviours. The HXR in panel (d) increases following helicon application, while in panel (e) the HXR signals decrease. The long-time scale behaviours though are the same in that the HXR growth is not as high as it would have been had helicon power not been turned on. This can be seen by the dashed lines in (d) and (e) of Fig \ref{fig:stdBttimetraces} which show the exponential growth rate before helicon being extrapolated to later times.
\par
\begin{figure}
    \centering
     \includegraphics[width=\columnwidth]{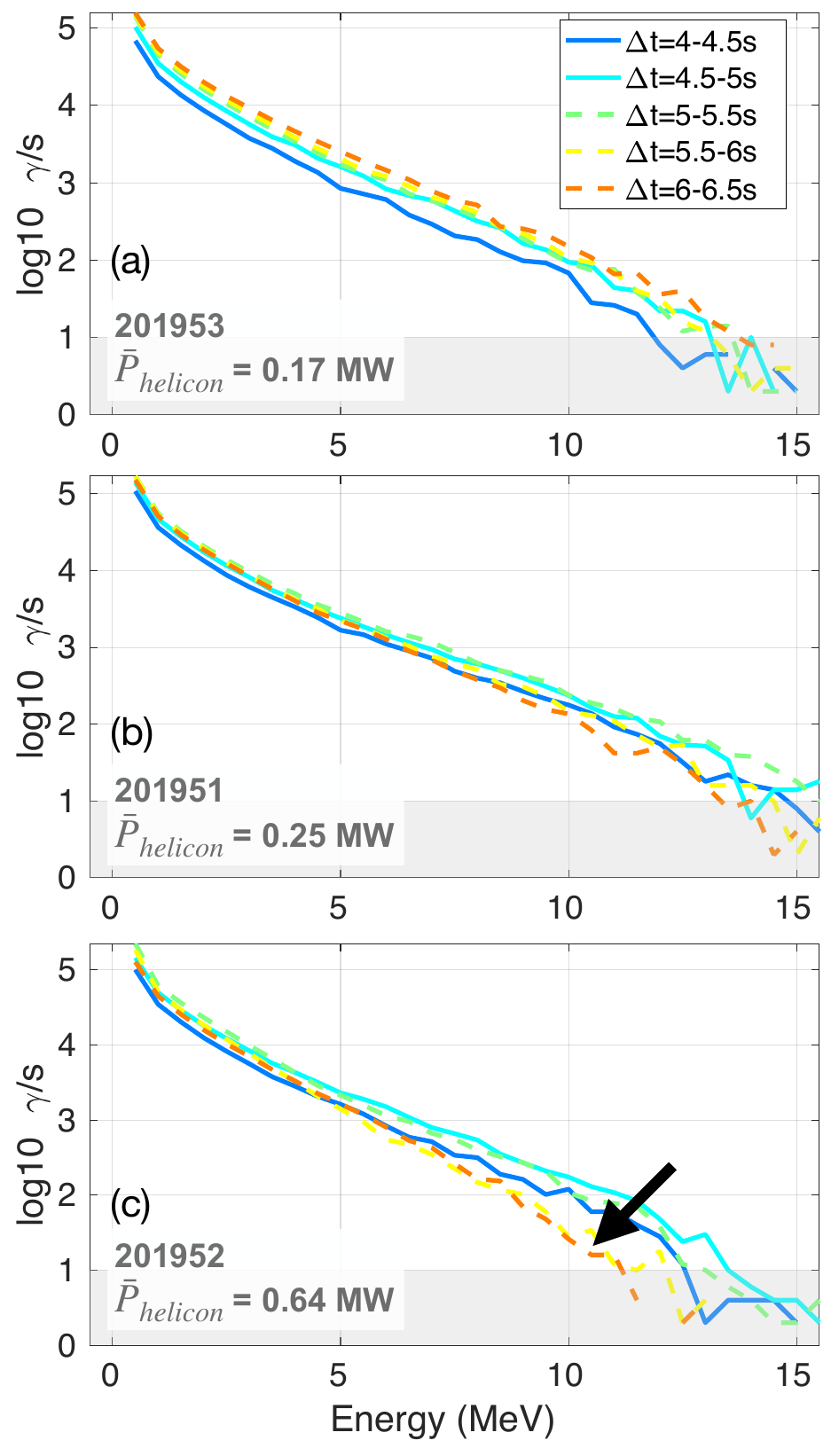}
    \caption{Effects of different average helicon power on the GRI-measured HXR spectrum in non-ideal configuration shots. All shots have helicon turn on at t=\SI{5}{s}. The average coupled helicon power \textit{during on pulses} for each shot is shown in each subplot. Note 201951 deposits less energy over the second or so of helicon whilst having a higher average coupled power during on pulses. The black arrow in panel (c) points in the direction the energy spectra are changing as time progresses. Time traces for these shots can be seen in Fig \ref{fig:stdBttimetraces}.}
    \label{fig:gripowerscan}
\end{figure}

For optimal helicon antenna operation, it was often desirable to set the helicon waveform to have the maximum power possible, and vary the timing, i.e., whether to have pulses and of what length, or whether to try and run continuous operation. It was challenging to cleanly launch the same waveform at different powers. But the three non-ideal configuration shots offer a glimpse at the effects of different average helicon powers on the RE population, since the three shots have very different average powers measured over the one second or so following the start of helicon power. Shown in Fig \ref{fig:gripowerscan} are the GRI-measured HXR spectra for these three shots. For the lowest average power case, shot 201953 with $\bar{P} \approx \SI{0.17}{MW}$, there is little change to the HXR spectrum following helicon application. As the power is increased to $\bar{P} \approx \SI{0.24}{MW}$ in shot 201952, there is a small drop in the HXR spectrum at high energies. And finally at the highest power shot there is a strong reduction in the HXR spectrum at energies above \SI{7}{MeV} or so, similar to the ideal configuration shots. This again suggests there is a power threshold for significant RE scattering via the normal Doppler resonance, though more shots would be needed to accurately determine this threshold and its dependence on key plasma parameters.
\par

\section{Discussion}
\begin{figure}
    \centering
    \includegraphics[width=\columnwidth]{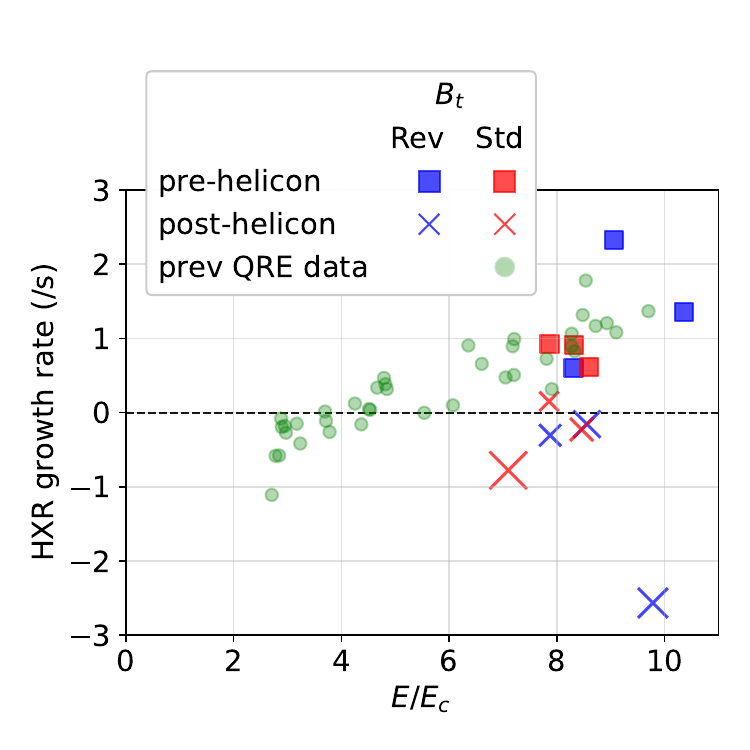}
    \caption{The HXR growth rate as a function of normalized electric field for QRE plasmas both with and without helicon waves in both configurations. The HXR, as measured by a plastic scintillator, is taken to scale as $\propto \exp\left( \gamma t \right)$, where $\gamma$ is the growth rate. The growth rate is calculated by fitting an exponential to the HXR signal over a time window from the start of helicon till \SI{1}{s} later or until a locked mode occurs. Reversed $ B_T$ shots are the ideal configuration and standard $ B_T$ shots are the non-ideal configuration. The green dots are from Ref \cite{paz-soldangrowthdecayrunaway2014} and do not include helicon waves. The different sizes of the `X' markers for post-helicon data points is to show the relative integrated helicon powers in those shots, i.e., larger markers have larger integrated helicon powers.}
    \label{fig:growthratevsEEc}
\end{figure}

The ability of helicon waves to resonantly scatter REs via the normal Doppler resonance in both ideal and non-ideal configurations can be understood by considering how the HXR growth rate depends on $E/E_c$. We show this dependence in Fig \ref{fig:growthratevsEEc}, where the green dots are from other QRE experiments that do not include helicon waves \cite{paz-soldangrowthdecayrunaway2014}. The new data, shown in blue and red for ideal and non-ideal configurations respectively, show that helicon waves are able to significantly decrease the HXR growth rate whilst not significantly changing the normalized electric field. This is consistent with the idea that helicon waves scatter REs via the normal Doppler as opposed to changing the background parameters like density or temperature significantly.

\subsection{Challenges involved in the application of this technique to Post-Disruption REs}
As discussed in the End Matter of Ref \cite{choudhuryFirstDemonstrationResonant2026}, the success of this technique in the QRE scenario in DIII-D is unlikely to extend to post-disruption RE beams without any significant modifications. The primary reason is that the cold plasmas have very high collisional damping, a detailed discussion of which is given in Appendix \ref{subsec:wavedamping}.
\par
To simulate how far the fast waves could travel, GENRAY simulations were performed for a post-disruption \SI{2.14}{T} on-axis field RE beam plasma. The magnetic equilibrium from shot 200809 at t=\SI{0.5}{s} was used alongside its actual measured $n_e$ density profile. A flat temperature profile was assumed. The plasma was also taken to be pure deuterium with $ Z_{\mathrm{eff}} = 1$. The waves were launched with $f= \SI{476}{MHz}$ and $n_\parallel = 4$ from a normalized minor radius of 0.02 on the \textit{inboard} side. The reason the wave had to be launched from the inboard side is that generally RE beams on DIII-D limit on the center-post and their cross sectional area does not fill the whole poloidal cross section. This means there is an outboard gap of roughly \SIrange{30}{50}{cm} between the last closed flux surface and the antenna. This is almost certainly too large to allow the fast wave to tunnel through the vacuum gap and reach the plasma. The wave is therefore required to be launched from the inboard side. For future devices, a location where RE beams are modeled to limit against could be chosen as a suitable location for a helicon antenna. The reason the wave was launched just inside the last-closed flux surface is that GENRAY simulations do not include the region beyond the last-closed flux surface, and so waves must be launched from within the closed-field line region.
\par
\begin{figure*}
    \centering
    \includegraphics[width=\textwidth]{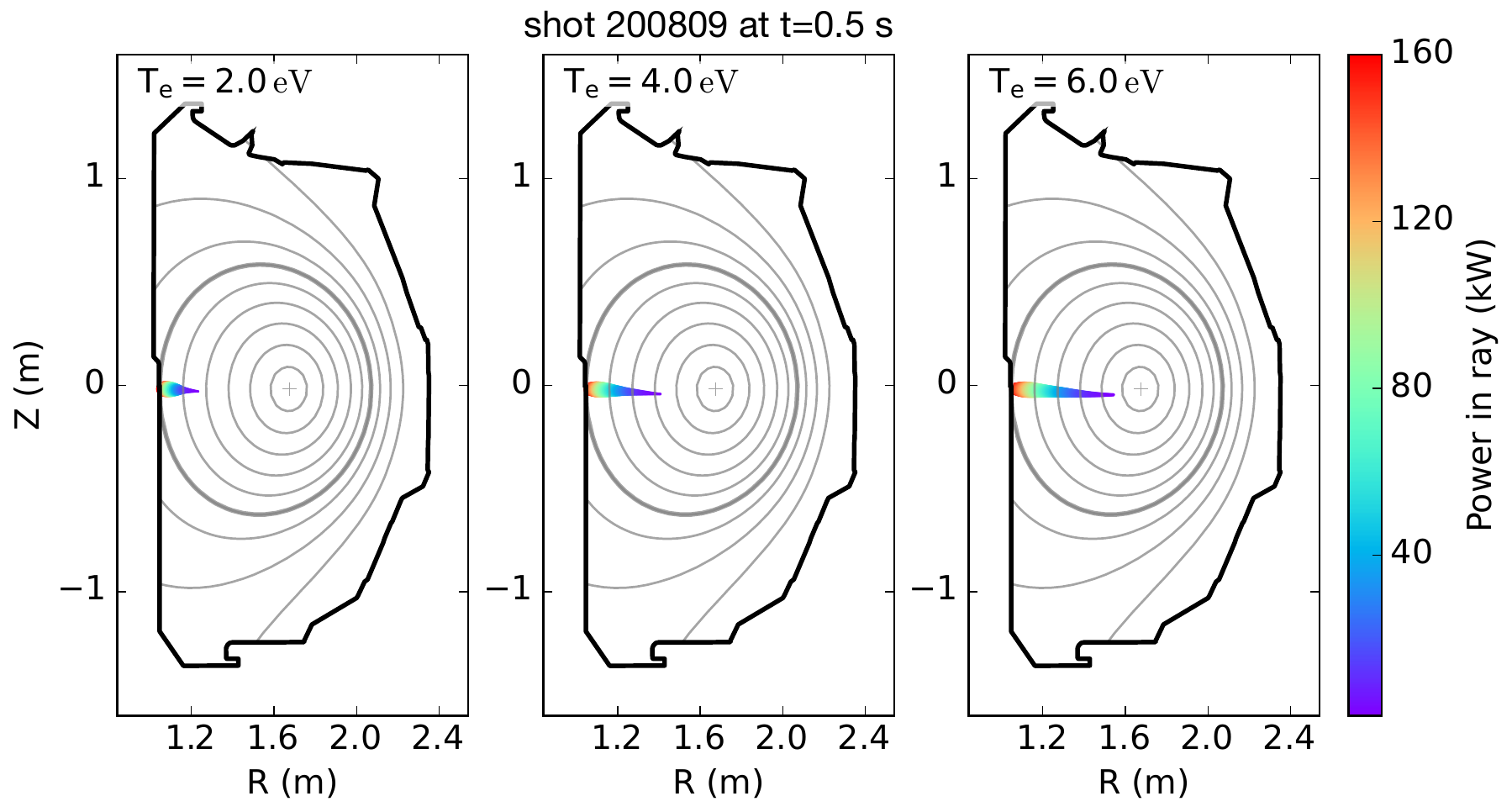}
    \caption{GENRAY simulations of inboard $n_\parallel = 4$ fast wave propagation in a post-disruption RE beam plasma for different temperature plasmas at a fixed density profile. The equilibrium used is from shot 200809 at t=\SI{0.5}{s}. The wave is launched from a normalized minor radius of 0.02 on the inboard side. The three different temperature plasmas shown are \SI{2}{eV} (left), \SI{4}{eV} (middle), and \SI{6}{eV} (right). The color of the ray traces indicates the wave power in kilowatts, see color-bar. The thickness of the ray is proportional to the rate of wave power loss to collisional damping.}
    \label{fig:post disruption genrays different temps}
\end{figure*}
\begin{figure}
    \centering    
    \includegraphics[width=\columnwidth]{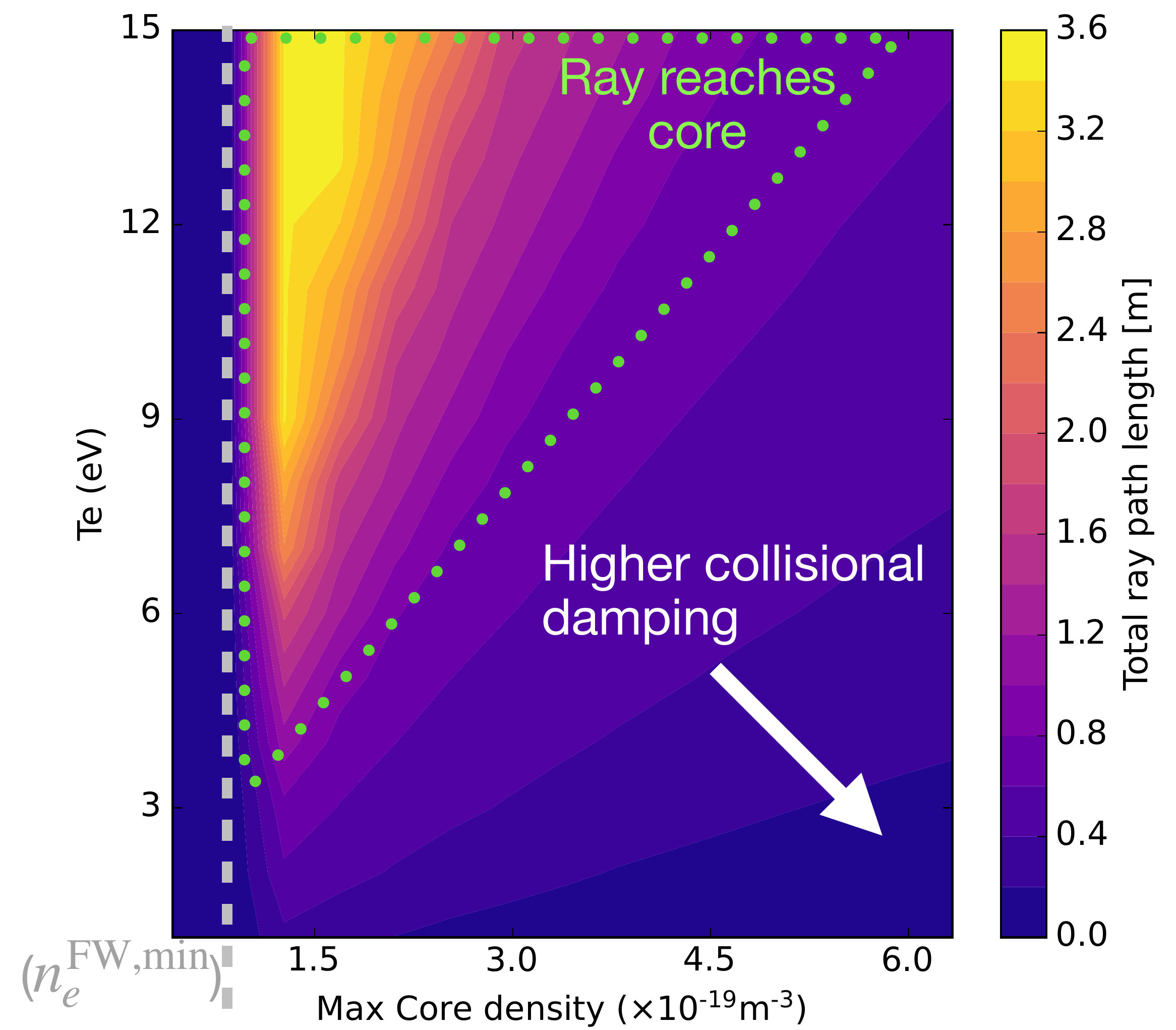}
    \caption{Temperature and density scan for fast-wave propagation in a post-disruption RE beam plasma. The white arrow illustrates the general direction of increased collisional damping as temperature decreases and density increases. The grey vertical line on the left shows roughly the core density below which the fast wave is not even able to propagate at all. Note since the density profile is not flat, and the wave is launched from a normalized minor radius of 0.02 on the inboard side, this grey line is meant to illustrate how low the core density can be dropped before the wave cannot propagate at all. The green dotted region shows the region of parameter space where the wave makes it into the core.}
    \label{fig:Tevsdensityscan postdisruption}
\end{figure}

The poloidal projection of several ray trajectories is shown in Fig \ref{fig:post disruption genrays different temps} for different temperature plasmas at a fixed density profile. The outboard gap can be seen to be prohibitively large, and so the wave is launched from the inboard side. The same density profile is used in all three cases with a maximum core density of $n_e \approx \SI{4.2e19}{m^{-3}}$. 
A scan of temperature-density parameter space is shown in Fig \ref{fig:Tevsdensityscan postdisruption}. The same equilibrium that was used in the poloidal projections in Fig \ref{fig:post disruption genrays different temps} is used for this parameter scan too.
\par
In summary, the challenges faced by the application of this technique to post-disruption RE beams on DIII-D are as follows: First, in cases where RE beams limit on the center post, the large outboard gap between the last closed flux surface and the antenna means the wave must be launched from the inboard side, where the antenna-plasma gap is much smaller. Second, the density and temperature of the background plasma have a very narrow operating window in which the fast wave can propagate into the core of the plasma with sufficient power to scatter REs. If the density is too low, the wave cannot propagate at all owing to the fast wave lower density cutoff. If the density is too high, or the temperature too low, collisional damping is so strong that the wave loses all its energy before reaching the core. And finally, impurities, which are certain to be present in significant quantities in post-disruption plasmas, will further complicate this picture and change the cut-off density and increase collisional damping by increasing $Z_{\mathrm{eff}}$, see Eq \ref{eq:collisionaldampingsimplified}.
\par
Regarding the application of this technique to future high-current devices like SPARC or ITER, integrated modelling will have to be performed to determine whether helicon scattering of post-disruption high-energy REs would be practical. Though modelling of the background plasma during the RE plateau has been done in the absence of helicon waves \cite{hollmannTrendsRunawayElectron2023}, there is scope for further modelling since the waves parallel refractive index and frequency and antenna position are free parameters that can be optimized for RE scattering. For reference, the DIII-D setup used in this experiment was optimized for high-beta current drive.

\section{Conclusion}
This paper builds upon and adds details to initial experimental results shown in Ref \cite{choudhuryFirstDemonstrationResonant2026} that demonstrated the ability of helicon waves to resonantly scatter REs via the normal Doppler resonance in DIII-D QRE plasmas. In this paper we showed that care must be taken in the \textit{direction} of the launched waves, since launching waves in the direction that favours the anomalous Doppler resonance has been found to enhance the RE population instead of reducing it. Further, this technique of RE scattering was not found to be particularly sensitive to whether the background total magnetic field aligned perfectly with the antenna's Faraday screen elements, nor whether the outboard gap was minimized. In such non-ideal configuration shots, a power threshold has been observed below which there is no significant scattering. Finally, as mentioned in Ref \cite{choudhuryFirstDemonstrationResonant2026}, the application of this technique to post-disruption RE beams is likely to be very challenging due to the high collisional damping in cold post-disruption plasmas and the large outboard gaps that are often present in DIII-D RE beam scenarios.

\ack{
The authors wish to thank Bob Pinsker for fruitful discussions regarding helicon wave physics.
}

\funding{
This material is based upon work supported by the U.S. Department of Energy, Office of Science, Office of Fusion Energy Sciences, using the DIII-D National Fusion Facility, a DOE Office of Science user facility, under Awards 
DE-FC02-04ER54698, 
DE-SC0022270, 
DE-SC0021622, 
DE-SC0020337, 
DE-AC05-00OR22725, 
and  
DE-FG02-07ER54917. 
\par
We thank the U.S. Department of Energy Office of Fusion Energy Sciences through the Basic Plasma Science Program at Los Alamos National Laboratory (LANL) under contract No. 89233218CNA000001. This research used resources of the National Energy Research Scientific Computing Center, a DOE Office of Science User Facility supported by the Office of Science of the U.S. Department of Energy under Contract No. DE-AC02-05CH11231 using
NERSC award FES-ERCAP0032298 and LANL Institutional Computing Program, which is supported by the U.S. Department of Energy National Nuclear Security Administration under Contract No. 89233218CNA000001.
\par
Disclaimer: This report was prepared as an account of work sponsored by an agency of the United States Government. Neither the United States Government nor any agency thereof, nor any of their employees, makes any warranty, express or implied, or assumes any legal liability or responsibility for the accuracy, completeness, or usefulness of any information, apparatus, product, or process disclosed, or represents that its use would not infringe privately owned rights. Reference herein to any specific commercial product, process, or service by trade name, trademark, manufacturer, or otherwise does not necessarily constitute or imply its endorsement, recommendation, or favoring by the United States Government or any agency thereof. The views and opinions of authors expressed herein do not necessarily state or reflect those of the United States Government or any agency thereof.
}

\roles{
H. Choudhury - Formal Analysis, Investigation, Methodology, Writing. 
A. Battey, J. Lestz, C. Paz-Soldan - Supervision, Methodology, Investigation.
D. Spong, W. Heidbrink - Conceptualization. 
X. Tang, Q. Zhang, and Y. Zhang - Methodology.
The remaining authors all contributed equally to investigation and data curation. 
}

\data{
The data that support the findings of
this article are available from the DIII-D user facility. 
}
%






\appendix

\section{$n_\parallel$ power spectrum used in GENRAY simulations}
\label{app_sec:npara_power_spectrum}
\begin{figure}
    \centering
    \includegraphics[width=\columnwidth]{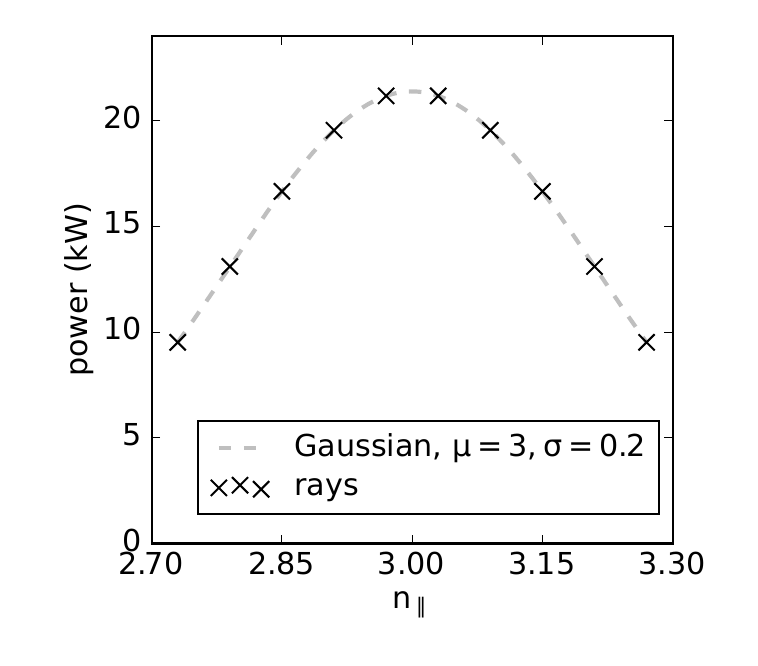}
    \caption{The powers of the 10 rays simulated in GENRAY in Figs \ref{fig:nparandnperp_vs_distancealongray} and \ref{fig:poloidal_trajectory_genray}. They are drawn from a gaussian centred on 3 with standard deviation of 0.21.}
    \label{fig:npara_power_spectrum}
\end{figure}

The power spectrum, P($n_\parallel$), for the GENRAY simulations presented in Figs \ref{fig:nparandnperp_vs_distancealongray} and \ref{fig:poloidal_trajectory_genray} is a gaussian with mean of 3 and standard deviation of 0.2, and the points were taken evenly spaced from the $n_\parallel$ range 3 - 0.2$\sqrt{2}$ to 3 + 0.2$\sqrt{2}$, as shown in Fig \ref{fig:npara_power_spectrum}

\section{Wave Damping and Reflection}
\label{subsec:wavedamping}
Radio-frequency waves in plasma can experience either collisional or collisionless damping. 
\par
As shown in Ref \cite{aleynikovstabilityanalysisrunawaydriven2015}, taking $\omega \gg \nu_{ee}$, collisional damping adds a small anti-hermitian component to the dielectric tensor:
\begin{equation}
    \uuline{\epsilon} = \uuline{\epsilon}^H + \uuline{\epsilon}^A,
\end{equation}
where the hermitian component is the usual cold-plasma dielectric tensor, which using the usual Stix symbols \cite{stixwavesplasmas1992}, is given by
\begin{equation}
    \uuline{\epsilon}^H = \begin{pmatrix}
    S & -iD & 0 \\
    iD & S & 0 \\
    0 & 0 & P
    \end{pmatrix}.
\end{equation}
The small anti-hermitian component is given by 
\begin{equation}
    \uuline{\epsilon}^A = \begin{pmatrix}
    S^A & -iD^A & 0 \\
    iD^A & S^A & 0 \\
    0 & 0 & P^A 
    \end{pmatrix},
\end{equation}
where
\begin{equation}
\begin{aligned}
    S^A & =i \frac{\nu_{\mathrm{ei}}}{\omega} \frac{\omega_{\mathrm{pe}}^2 \omega^2+\omega_{\mathrm{ce}}^2 \omega_{\mathrm{pe}}^2}{\left(\omega^2-\omega_{\mathrm{ce}}^2\right)^2} \\
    D^A & = 2 i \frac{\nu_{\mathrm{ei}}}{\omega} \frac{\omega_{\mathrm{pe}}^2 \omega\left|\omega_{\mathrm{ce}}\right|}{\left(\omega^2-\omega_{\mathrm{ce}}^2\right)^2} \\
    P^A & =i \frac{\nu_{\mathrm{ei}}}{\omega} \frac{\omega_{\mathrm{pe}}^2}{\omega^2}
\end{aligned}
\end{equation}
where 
\begin{equation}
    \nu_{e i}=\frac{4 \sqrt{2 \pi} e^4 \ln \Lambda}{3 m^{1 / 2} T_e^{3 / 2}} \sum_Z Z^2 n_Z
    \label{eq:electron_ion_collision_frequency}
\end{equation}
is the electron-ion collision frequency. Using this modified dielectric tensor, the wave damping rate can be shown as \cite{aleynikovstabilityanalysisrunawaydriven2015}
\begin{equation}
    \Gamma_\nu
    = -i \omega^2 \frac{ \vec{E}^\dagger \uuline{{\epsilon}} \vec{E}}{\vec{E}^\dagger (\frac{\partial}{\partial \omega} \omega^2) \vec{E}}
    \label{eq:collisionaldamping}
\end{equation}
which can be simplified to 
\begin{equation}
    \Gamma_\nu
    = \nu_{ e i} \frac{ \vec{E}^\dagger \frac{\partial}{\partial \omega} \left( \omega (\uuline{\epsilon}^H - \mathbf{I})  \right) \vec{E}}{\vec{E}^\dagger (\frac{\partial}{\partial \omega} (\omega^2 \uuline{{\epsilon}}^H)) \vec{E}}
    \label{eq:collisionaldampingsimplified}
\end{equation}
for the case of high-frequency electron waves that do not involve ion motion. For a single species hydrogenic plasma, the temperature dependence dominates the collisional damping rate, $\Gamma_\nu \propto T_e^{-3/2}$ \cite{casolariwhistlerwavedestabilization2022}. For the QRE plasmas shown in this paper, typically of core temperatures of at least \SI{2}{keV}, the collisional damping rate is negligible.
\par
Helicon waves also experience parallel collisionless damping, also known as Landau damping, which in a warm plasma of temperature $T_e$ and density $n_e$ scales as \cite{praterapplicationveryhigh2014,compernollehighpowerheliconprogram2021}
\begin{equation}
    k_{\perp \mathrm{i}} \propto k_{\perp} \beta_{\mathrm{e}} \xi_{\mathrm{e}} \mathrm{e}^{-\xi_{\mathrm{e}}^2},
    \label{eq:landaudampingkperpscaling}
\end{equation}
where $k_{\perp \mathrm{i}}$ is the spatial decay length scale, $\beta_e = n_e k_B T_e / (B^2/2 \mu_0)$ is the ratio of electron thermal pressure to magnetic pressure and $\xi_e = \frac{v_{ph, \parallel}}{v_{\mathrm{th}}} = \omega / (k_\parallel v_{\mathrm{th}})$ is the ratio of parallel wave phase velocity to electron thermal velocity. QRE plasmas in general are low $\beta_e$ plasmas since for $n_e \sim \SI{5e18}{m^{-3}}$, $T_e \sim \SI{2}{keV}$ and $B \sim \SI{1.4}{T}$, $\beta_e \approx 0.2\%$ (c.f. the high-beta current-drive scenario for which the antenna was designed, where $\beta_e \geq 2\%$ \cite{praterapplicationveryhigh2014}). 
\par
\par
QRE plasmas therefore exist in an interesting high-temperature low-density regime where \textit{both} collisional and collisionless damping are weak. The waves will therefore very probably undergo many passes and reflections inside the plasma unless they resonate with REs and lose energy to them. This suggests that even though the waves are launched with a well-defined $n_{\parallel, antenna}$ spectrum, after many passes the waves will have a broadened $n_\parallel$ spectrum and potentially be stochasticized in phase space. This could be beneficial for resonant wave-particle interactions since a broader $n_\parallel$ spectrum means a broader range of RE energies can satisfy the resonance condition. Another possibility is that the fast wave will convert to the slow wave. As can be seen from Eq \ref{eq:landaudampingkperpscaling}, slow waves undergo much stronger Landau damping owing to their much larger $k_\perp$ values. For example, solving the cold plasma dispersion with $n_\parallel = 3, \omega= 2 \pi \times \SI{476}{MHz}$, $B = \SI{1.4}{T}$ and $n_e = \SI{5e18}{m^{-3}}$, the two solutions have $k_\perp = \SI{69}{m^{-1}}$ for the fast wave and $k_\perp = \SI{1360}{m^{-1}}$ for the slow wave ($\omega/c\approx \SI{10}{m^{-1}}$ for these waves). This large difference in $k_\perp$ means that if mode conversion from fast to slow wave occurs, the wave energy will be rapidly damped on thermal electrons.

\section{Pitch-angle of resonant REs}
\label{app:pitch-angle-derivation}
Since the parallel velocity $v_{\parallel}$ of runaway electrons (REs) is typically
much larger than their perpendicular velocity $v_{\perp}$, we can take $\mathbf{k}\cdot\mathbf{v} \approx k_{\parallel} v_{\parallel}$.
Introducing the pitch-angle cosine $\xi = v_{\parallel}/v$, the resonance
condition becomes
\begin{equation}
\omega - k_{\parallel} v\,\xi = \frac{n\,\Omega_{ce}}{\gamma}.
\end{equation}
Solving for $\xi$ gives
\begin{equation}
\xi = \frac{\omega\,\gamma - n\,\Omega_{ce}}
{k_{\parallel}\,v \gamma}.
\label{xi_basic}
\end{equation}

To evaluate $\xi$ for a given kinetic energy $E_k$, we first relate $E_k$ to
the Lorentz factor. The total energy satisfies
\begin{equation}
E_{\mathrm{tot}} = \gamma m_e c^2 = m_e c^2 + E_k,
\end{equation}
so that
\begin{equation}
\gamma(E_k) = 1 + \frac{E_k}{m_e c^2}.
\end{equation}
The corresponding relativistic momentum follows from the
energy--momentum relation $\gamma^2 = 1 + p^2/(m_e^2 c^2)$:
\begin{equation}
p(E_k) = m_e c\,\sqrt{\gamma(E_k)^2 - 1}.
\end{equation}
Using $p = \gamma m_e v$, the speed can be written as
\begin{equation}
v(E_k) = \frac{p(E_k)}{\gamma(E_k)\,m_e}
       = c\,\sqrt{1 - \frac{1}{\gamma(E_k)^2}}.
\end{equation}

Substituting these expressions into the resonance condition gives an
explicit energy dependence for the resonant pitch angle. In the form of
Eq.\,\ref{xi_basic}, this can be written as
\begin{equation}
\xi(E_k) =
\frac{\gamma(E_k)\,\omega - n\,\Omega_{ce}}
{k_{\parallel}\,c\,\gamma(E_k)\,
 \sqrt{1 - 1/\gamma(E_k)^2}}.
\end{equation}
Alternatively, using $v = p/(\gamma m_e)$, Eq.\,\ref{xi_basic} is
equivalently
\begin{equation}
\xi(E_k) =
\frac{m_e\big[\gamma(E_k)\,\omega - n\,\Omega_{ce}\big]}
{k_{\parallel}\,p(E_k)},
\end{equation}
which is the form directly implemented in the numerical evaluation of
$\xi(E_k)$.

\bibliographystyle{iopart-num}

\bibliography{references}

\end{document}